\documentclass[onecolumn]{IEEEtran}

\usepackage{pgf,tikz}
\usetikzlibrary{arrows}
\usepackage[margin=0.7in]{geometry}
\usepackage{amssymb}
\usepackage{amsmath}
\usepackage{amsthm}
\usepackage{color}
\usepackage{hyperref}
\usepackage{mathtools}
\usepackage{multirow}
\usepackage{nicefrac}
\usepackage{hhline}
\usepackage{tabularx}
\usepackage[boxruled, linesnumbered]{algorithm2e}
\usepackage{diagbox}

\DeclarePairedDelimiter{\floor}{\lfloor}{\rfloor}
\DeclarePairedDelimiter{\bigfloor}{\Bigl\lfloor}{\Bigr\rfloor}

\newtheorem{theorem}{Theorem}
\newtheorem{definition}{Definition}

\newtheorem{lemma}{Lemma}
\newtheorem{corollary}{Corollary}

\newtheorem{example}{Example}

\newcommand{\bN}{\mathbb{N}}

\newcommand{\bR}{\mathbb{R}}

\newcommand{\bZ}{\mathbb{Z}}

\newcommand{\cA}{\mathcal{A}}

\newcommand{\cC}{\mathcal{C}}

\newcommand{\cF}{\mathcal{F}}

\newcommand{\cH}{\mathcal{H}}

\newcommand{\cP}{\mathcal{P}}

\newcommand{\cS}{\mathcal{S}}

\newcommand{\qnum}[2]{\left<#1,#2\right>}

\DeclareMathOperator{\len}{len}

\newcommand{\parenv}[1]{\left( #1 \right)}

\makeatletter
\def\namedlabel#1#2{\begingroup
	\def\@currentlabel{#2}%
	\label{#1}\endgroup
}
\makeatother

\title{Rank-Modulation Codes for DNA Storage}

\author{ \textbf{Netanel Raviv$^{\star,\dagger}$}, \textbf{Moshe
    Schwartz$^\dagger$}, and \textbf{Eitan
    Yaakobi$^\star$}\\ \IEEEauthorblockA{$^\star$Computer Science
    Department, Technion -- Israel Institute of Technology, Haifa
    3200003, Israel\\ $^\dagger$Electrical and Computer Engineering,
    Ben-Gurion University of the Negev, Beer Sheva 8410501,
    Israel\\ \textit{netanel.raviv@gmail.com, schwartz@ee.bgu.ac.il,
      yaakobi@cs.technion.ac.il}} \thanks{The material in this paper
    was presented in part at the IEEE International Symposium on
    Information Theory (ISIT 2017), Aachen, Germany, June 2017. This work was supported in part by the Israel Science Foundation under grant no.~130/14 and grant no.~1624/14.}  }

\begin{document}
\maketitle
\thispagestyle{empty}

\begin{abstract}
Synthesis of DNA molecules offers unprecedented advances in storage technology. Yet, the microscopic world in which these molecules reside induces error patterns that are fundamentally different from their digital counterparts. Hence, to maintain reliability in reading and writing, new coding schemes must be developed. 

In a reading technique called shotgun sequencing, a long DNA string is read in a sliding window fashion, and a profile vector is produced. It was recently suggested by Kiah et al. that such a vector can represent the permutation which is induced by its entries, and hence a rank-modulation scheme arises. Although this interpretation suggests high error tolerance, it is unclear which permutations are feasible, and how to produce a DNA string whose profile vector induces a given permutation. 

In this paper, by observing some necessary conditions, an upper bound for the number of feasible permutations is given. Further, a technique for deciding the feasibility of a permutation is devised. By using insights from this technique, an algorithm for producing a considerable number of feasible permutations is given, which applies to any alphabet size and any window length.
\end{abstract}


\hypersetup{hyperfootnotes=true}

\section{Introduction}\label{section:Introduction}
Coding for DNA storage devices has gained increasing attention lately, following a proof of concept by several promising prototypes~\cite{BorLopCarDouCezSeeStr16,ChuGaoKos12,GolBerCheDesLepSipBri13}. Due to the high cost and technical limitations of the synthesis (i.e., writing) process, these works focused on producing short strings, but as the cost of synthesis declines, longer strings can be produced. In turn, long strings are read more accurately by a technique called \emph{shotgun sequencing}~\cite{MotBreTse13}, in which short substrings are read separately and reassembled together to form the original string. 

The shotgun-sequencing technique has motivated the definition of the DNA storage channel~\cite{KiaPulMil16} (see Figure~\ref{figure:DNAstorageChannel}). In a channel with alphabet~$\Sigma$ of size~$q$, and window length~$\ell$, data is stored as a (possibly circular) string over~$\Sigma$; the output of the channel is a (possibly erroneous) \emph{histogram}, or a \emph{profile vector} with~$q^\ell$ entries, each containing the number of times that the corresponding $\ell$-substring was observed. Errors in this channel might occur as a result of substitutions in the synthesis or sequencing processes, or imperfect coverage of reads. 

In order to cope with different error patterns, several code constructions were recently suggested~\cite{KiaPulMil16}, however, much is yet to be done to obtain high error resilience and high rate. It was also suggested in~\cite[Sec.~VIII.B]{KiaPulMil16} to employ a rank-modulation scheme, which has the potential for coping with \emph{any} error pattern that does not revert the order among the entries of the profile vector.

\definecolor{cqcqcq}{rgb}{0.3,0.3,0.3}
\begin{figure}
\begin{center}
\begin{tikzpicture}[line cap=round,line join=round,>=triangle 45,x=1cm,y=1cm]
\clip(-1,2.4) rectangle (13.42296072507553,9.2);
\fill[color=cqcqcq,fill=cqcqcq,fill opacity=0.1] (5,8) -- (7,8) -- (7,9) -- (5,9) -- cycle;
\fill[color=cqcqcq,fill=cqcqcq,fill opacity=0.1] (0.75,4.85) -- (0.75,7.15) -- (11.25,7.15) -- (11.25,4.85) -- cycle;
\fill[color=cqcqcq,fill=cqcqcq,fill opacity=0.1] (0.75,2.5) -- (0.75,4) -- (11.25,4) -- (11.25,2.5) -- cycle;
\draw [color=cqcqcq,line width=1.5pt] (5,8)-- (7,8);
\draw [color=cqcqcq,line width=1.5pt] (7,8)-- (7,9);
\draw [color=cqcqcq,line width=1.5pt] (7,9)-- (5,9);
\draw [color=cqcqcq,line width=1.5pt] (5,9)-- (5,8);
\draw [color=cqcqcq,line width=1.5pt] (0.75,4.85)-- (0.75,7.15);
\draw [color=cqcqcq,line width=1.5pt] (0.75,7.15)-- (11.25,7.15);
\draw [color=cqcqcq,line width=1.5pt] (11.25,7.15)-- (11.25,4.85);
\draw [color=cqcqcq,line width=1.5pt] (11.25,4.85)-- (0.75,4.85);
\draw (6,5.35) node[anchor=center] {$(\texttt{AA}: 2,\texttt{AC}:4,\texttt{AG}:10,\texttt{CA}:6,\texttt{CC}:12,\texttt{CG}:14, \texttt{GA}:8,\texttt{GC}:16,\texttt{GG}:18)$};
\draw (6,6.15) node[anchor=center] {$\texttt{AGGGGGGGGGGCGCGCGCGCGCGCGAGAGAGAGCCCCCCCACACA-}$};
\draw (6,5.85) node[anchor=center] {$\texttt{-AGGGGGGGGGGCGCGCGCGCGCGCGAGAGAGAGCCCCCCCACACA}$};
\draw (6,8.5) node[anchor=center] {\textbf{Data}};
\draw [->] (6,8) -- (6,7.15);
\draw (6.25,7.575) node[anchor=west] {\small\textit{synthesis}};
\draw [color=cqcqcq,line width=1.5pt] (0.75,2.5)-- (0.75,4);
\draw [color=cqcqcq,line width=1.5pt] (0.75,4)-- (11.25,4);
\draw [color=cqcqcq,line width=1.5pt] (11.25,4)-- (11.25,2.5);
\draw [color=cqcqcq,line width=1.5pt] (11.25,2.5)-- (0.75,2.5);
\draw (6,3) node[anchor=center] {$(\texttt{AA}:2,\texttt{AC}:4,\texttt{AG}:10,\texttt{CA}:6,\texttt{CC}:\textcolor{red}{13},\texttt{CG}:14, \texttt{GA}:8,\texttt{GC}:16,\texttt{GG}:\textcolor{red}{17})$};
\draw [->] (6,4.85) -- (6,4);
\draw (6.25,4.425) node[anchor=west] {\small\textit{shotgun sequencing}};
\draw (6,6.65) node[anchor=center] {\textbf{Storage}};
\draw (6,3.5) node[anchor=center] {\textbf{Output}};
\end{tikzpicture}
\caption{An example of a transmission in the DNA storage channel with the parameters~$q=3$, $\ell=2$, and~$n=90$. The data is encoded into a circular DNA string using a process called \textit{synthesis}. The stored string is read using a process called \textit{shotgun sequencing}, whose output is a (possibly erroneous) profile vector.}\label{figure:DNAstorageChannel}
\end{center}
\end{figure}
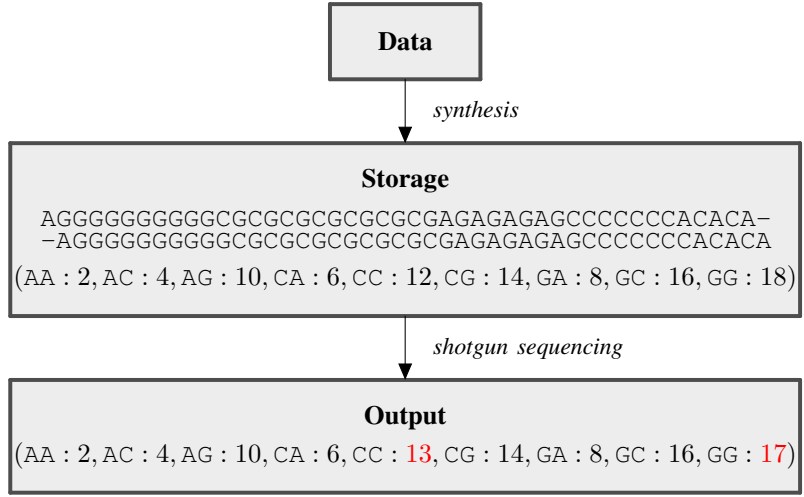

In this scheme, the absolute values of the profile vector are ignored, and only their ranks in relation to each other are considered. That is, a given profile vector represents the permutation which is induced by its entries. Clearly, to induce a permutation the entries of the profile vector must be distinct, which is a reasonable assumption for long strings and short window length. For example, the (actual) stored string in Figure~\ref{figure:DNAstorageChannel} represents the permutation
\begin{align}\label{equation:examplePi}
\pi=(\texttt{AA},\texttt{AC},\texttt{CA},\texttt{GA},\texttt{AG},\texttt{CC},\texttt{CG},\texttt{GC},\texttt{GG}),
\end{align}
since $\texttt{AA}$ has the lowest frequency, followed by
$\texttt{AC}$, and so on, until reaching $\texttt{GG}$ with the
highest frequency.  Furthermore, it is evident that the noisy output
in Figure~\ref{figure:DNAstorageChannel}, due perhaps to the
shotgun-sequencing process, still represents the same
permutation~$\pi$, and hence, this error pattern in the profile vector
is correctable by using a rank-modulation scheme.

A well known tool in the analysis of strings is the DeBruijn graph~$G_q^\ell$, whose set of nodes is~$\Sigma^\ell$, and two nodes are connected by a directed edge if the~$(\ell-1)$-suffix of the former is the~$(\ell-1)$-prefix of the latter. This graph is a useful tool in the analysis of the DNA storage channel since any string over~$\Sigma$ induces a path in the graph, and since a (normalized) profile vector can be seen as measure on its node set. Further, we may restrict our attention to closed strings only (i.e., strings that correspond to closed paths), since in our context, they are asymptotically equivalent to their ordinary counterparts.

Due to flow conservation constraints in the DeBruijn graph~\cite{JacKneSzp12}, it is evident that not every permutation is feasible, i.e., there exist permutations that are not induced by any profile vector, and hence by any string. Consequently, \cite{KiaPulMil16} suggested to disregard a certain subset~$S$ of the entries in the profile vector, encode any permutation on the complement of~$S$, and complete the entries of~$S$ to obtain flow conservation.

In this paper the feasibility question is studied in a more restrictive setting, where \emph{all} the entries of the profile vector are considered. By formulating several necessary conditions, an upper bound is given on the number of feasible permutations, out of all permutations on~$q^\ell$ elements. In addition, a linear-programming technique is devised to decide the feasibility of a given permutation. Using insights from this technique, an encoding algorithm for producing a large number of feasible permutations for any~$q$ and any~$\ell$ is given. Interestingly, some of the above results rely on an interpretation of the encoding process as a \emph{Markov chain} on the DeBruijn graph. 

Finally, this problem may also be seen in a more general setting, beyond any applications for DNA storage. For example, it may be seen as a highly-restrictive variant of \emph{constraint coding}, an area of coding theory that concerns the construction of strings with or without some prescribed substrings. This resemblance is apparent by observing that in our setting, \emph{every} $\ell$-substring is required to be more or less frequent than any other $\ell$-substring.

The paper is organized as follows. In Section~\ref{section:preliminaries} we provide notation as well as formal definitions which are used throughout the paper. In Section~\ref{section:upperBound} we prove an upper bound on the number of feasible permutations. An efficient algorithm for deciding whether a permutation is feasible is described in Section~\ref{section:polynomialAlgorithm}. We turn in Section~\ref{section:encoding} to devise a construction for a large set of feasible permutations, thus also providing a lower bound on their total number. An important parameter of interest is the length of the shortest string for a given feasible permutation. An upper bound on this parameter is given in Section~\ref{section:length}. The paper is concluded in Section~\ref{section:discussion} with a short discussion.

\section{Preliminaries}\label{section:preliminaries}
For an alphabet~$\Sigma\triangleq\{\sigma_1,\ldots,\sigma_q\}$ and a window size~$\ell\in\bN$, let~$G_q^\ell$ be the DeBruijn graph of order~$\ell$ over~$\Sigma$. That is, the node set~$V(G_q^\ell)$ is~$\Sigma^\ell$, and for any two nodes~$u$ and~$v$ in~$V(G_q^\ell)$, the edge set~$E(G_q^\ell)$ contains~$(u,v)$ if the~$(\ell-1)$-suffix of~$u$ equals the~$(\ell-1)$-prefix of~$v$. An edge~$(u,v)$ is labeled by a string~$w\in\Sigma^{\ell+1}$ whose~$\ell$-prefix is~$u$ and whose~$\ell$-suffix is~$v$.

For a string~$x\in\Sigma^n$, $x=x_0 x_1\dots x_{n-1}$, with $x_i\in\Sigma$, the $\ell$-profile vector~$p_x\in\bZ^{q^\ell}$ (or simply, the profile vector) is a vector with $q^\ell$ non-negative integer entries~$(p_x(w))_{w\in\Sigma^\ell}$, each of which contains the number of occurrences of the corresponding $\ell$-substring in~$x$. That is, $p_x(w)=\left|\{0\leq i\leq n-1 \mid x_i,\ldots,x_{i+\ell-1} =w\}\right|$ for all~$w\in\Sigma^\ell$, where indices are taken modulo~$|x|=n$. The entries of~$p_x$ may be identified by either the set of nodes of~$G_q^\ell$ or the set of edges of~$G_q^{\ell-1}$, and the subscript~$x$ is omitted if clear from context.

For a given set~$A$, let~$S_A$ be the \textit{set of permutations on~$A$}, i.e., the set of vectors in~$A^{|A|}$ in which every element of~$A$ appears exactly once. For brevity, let~$S_{q,\ell} \triangleq S_{\Sigma^\ell}$, and~$S_i\triangleq S_{[i]}$ for any positive integer~$i$, where~$[i]\triangleq \{1,2,\ldots,i\}$.
Let~$\pi\in S_{q,\ell}$ and~$p\in\bR^{q^\ell}$, where the coordinates of~$\bR^{q^\ell}$ are identified by the elements of~$\Sigma^{\ell}$, ordered lexicographically. We say that~$p$ \textit{satisfies}~$\pi$, and denote~$p\vDash\pi$, if the entries of~$p$ are distinct, and their ascending order matches~$\pi$, i.e., $p(w)<p(w')$ if and only if $\pi(w)<\pi(w')$, for all $w,w'\in\Sigma^\ell$.
Similarly, for a string~$x\in\Sigma^*$ (where~$\Sigma^*$ is the set of all closed strings) and a permutation~$\pi\in S_{q,\ell}$ we say that~$x$ satisfies~$\pi$, and denote~$x\vDash\pi$, if $p_x\vDash\pi$, i.e., if the profile vector of~$x$ satisfies~$\pi$.

For example, in Figure~\ref{figure:DNAstorageChannel}, the vector at the output of the channel is $p\triangleq(2,4,10,6,13,14,8,16,17)\in\bR^{9}$; and since the entries of~$\bR^9$ are indexed by the elements of~$\{A,C,G\}^2$, ordered lexicographically, it follows that~$p\vDash\pi$ for the permutation~$\pi$ that was given in~\eqref{equation:examplePi}. Consequently, the string~$x$ in the \textbf{Storage} phase of Figure~\ref{figure:DNAstorageChannel} satisfies~$\pi$.

A permutation~$\pi\in S_{q,\ell}$ is called \textit{feasible} if there exists a string~$x\in\Sigma^*$ such that~$x\vDash \pi$, and infeasible otherwise. Similarly, for any subset~$U\subseteq \Sigma^\ell$, a permutation on~$U$ is called \textit{infeasible} if it cannot be extended to a feasible permutation in~$S_{q,\ell}$. Clearly, only vectors with distinct entries can satisfy a permutation, and hence, not every string satisfies a permutation.

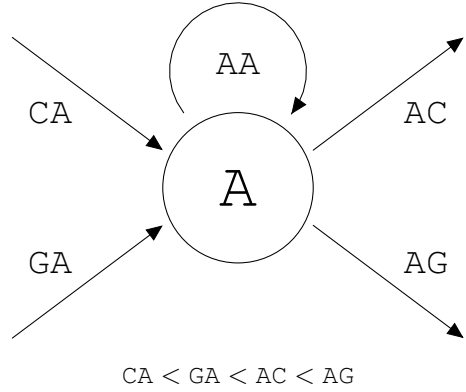
\begin{figure}
	\begin{center}
\begin{tikzpicture}[line cap=round,line join=round,>=triangle 45,x=1cm,y=1cm]
\clip(-4,-3) rectangle (4,3);
\draw(0,0) circle (1cm);
\draw [shift={(0,1.5513658545478795)}] plot[domain=-0.6778745994889395:3.7960190594317504,variable=\t]({1*0.9101469697935934*cos(\t r)+0*0.9101469697935934*sin(\t r)},{0*0.9101469697935934*cos(\t r)+1*0.9101469697935934*sin(\t r)});
\draw [->] (-3,2) -- (-1,0.5);
\draw [->] (-3,-2) -- (-1,-0.5);
\draw [->] (1, 0.5) -- (3,2);
\draw [->] (1,-0.5) -- (3,-2);
\draw (0,0) node[anchor=center] {\Huge$\texttt{A}$};
\draw (0,1.65) node[anchor=center] {\Large$\texttt{AA}$};
\draw (-2.5, 1) node[anchor=center] {\Large$\texttt{CA}$};
\draw (-2.5,-1) node[anchor=center] {\Large$\texttt{GA}$};
\draw ( 2.5, 1) node[anchor=center] {\Large$\texttt{AC}$};
\draw ( 2.5,-1) node[anchor=center] {\Large$\texttt{AG}$};
\draw [->] (0.7720726993740048,1.0694205739342004) -- (0.7089202084807104,0.980577238130097);
\draw (0,-2.5) node[anchor=center] {$\texttt{CA}<\texttt{GA}<\texttt{AC}<\texttt{AG}$};
\end{tikzpicture}
		\caption{An infeasible permutation. Since only closed strings are considered, any profile vector~$p$ must satisfy \textit{flow conservation constraints} at any node of~$G_q^{\ell-1}$. For example, if~$q=3$ and~$\ell=2$, the constraint that corresponds to the node~$\texttt{A}$ is~$p(\texttt{AA})+p(\texttt{CA})+p(\texttt{GA})=p(\texttt{AA})+p(\texttt{AC})+p(\texttt{AG})$. Hence, any permutation in which $(\texttt{CA},\texttt{GA},\texttt{AC},\texttt{AG})$ can be attained by deletion of entries is infeasible.}\label{figure:exampleInfeasiblePerm}
	\end{center}
\end{figure}

Another constraint on feasibility is \emph{flow conservation}. For every $\ell \geq 2$, and any $x\in\Sigma^*$, we must have
\[ \sum_{\sigma\in\Sigma} p_x(\sigma w) = \sum_{\sigma\in\Sigma} p_x(w\sigma),\]
for all $w\in\Sigma^{\ell-1}$. We call these the flow-conservation constraints, which easily follow by noting that each side of the equation equals the number of occurrences of~$w$ in~$x$. Another view of these constraints follows by noting that any string $x\in\Sigma^*$ may be scanned using a sliding window of length $\ell$, inducing cycle in the DeBruijn graph $G_q^{\ell-1}$. The flow-conservation constraints simply state that, along the cycle, the number of times we enter vertex $w\in\Sigma^{\ell-1}$ is $\sum_{\sigma\in\Sigma} p_x(\sigma w)$, must equal the number of times we exit this vertex, i.e., $\sum_{\sigma\in\Sigma} p_x(w\sigma)$. Due to flow conservation constraints in the DeBruijn graph, some permutations are infeasible, as illustrated in Figure~\ref{figure:exampleInfeasiblePerm}.

We note that, given a profile vector $p$ with flow conservation of order $\ell$, there exists a string $x\in\Sigma^*$ whose profile vector is $p$ provided another condition is met: Construct the DeBruijn graph $G_q^{\ell-1}$, and remove all edges $w\in\Sigma^\ell$ such that $p(w)=0$. Then remove all isolated vertices (i.e., vertices with no incoming edges and no outgoing edges). If the resulting graph is strongly connected, such a string $x$ exists. To see that, take $p(w)$ parallel copies of each edge $w$. Then, each vertex has in-degree that equals its out-degree (due to flow conservation of $p$), and since the graph is strongly connected, there exists an Eulerian cycle. The string $x$ associated with the Eulerian cycle (i.e., whose sequence of sliding windows of length $\ell$ equals the sequence of edges in the cycle) has a profile vector $p$. 

Given a (flow conserving) profile vector~$p$ such that~$p\vDash \pi$ for some permutation~$\pi$, the above implies a deterministic algorithm for generating a string~$x$ such that~$x\vDash \pi$. Alternatively, given any vector~$r\in\bR^{q^\ell}$ such that~$r\vDash\pi$ for some~$\pi$, it is possible to produce the corresponding string~$x$ by either turning it to an integer vector\footnote{This is possible by finding a close enough rational approximation, and multiplying by the least common multiple of its entries' denominators. Alternatively, one may multiply it by a large enough constant such that the absolute difference between any two distinct entries is at least $3$, and apply the algorithm of~\cite[Thm.~39]{EliMeySch16}. } which satisfies the same permutation and repeating the above, or by the following randomized algorithm.
	
Given such a vector~$r$, find~$\alpha$ and~$\beta$ in~$\bR$ such that~$s\triangleq\alpha r+\beta\mathbf{1}$ is a positive vector whose sum of entries is~$1$ (which clearly satisfies the same permutation as~$r$), and define~$M_s\in \bR^{q^\ell\times q^\ell}$ such that
\begin{align}\label{equation:TransitionMatrix}
(M_s)_{a,b}&=\begin{cases}
\frac{s(v\sigma)}{\sum_{\tau\in\Sigma}s(v\tau)} & \mbox{if }(a,b)\mbox{ is an edge and }b=v\sigma,\\
0 & \mbox{otherwise},
\end{cases}.
\end{align}
It is proved in Lemma~\ref{lemma:TransitionMatrix} in Appendix~\ref{appendix:OmittedProofs} that~$M_s$ is a transition matrix of a Markov chain on~$G_q^\ell$, and its stationary distribution is~$s$. Hence, it follows from the law of large numbers for Markov chains that following this chain for long enough produces a string whose profile vector satisfies the same permutation as~$s$ and~$r$.

The main goals of this paper are to characterize, bound, and construct sets of feasible permutations in~$S_{q,\ell}$. To this end, given~$q$ and~$\ell$, let 
\begin{align}\label{equation:FqlRql}
\nonumber\cF_{q,\ell}&\triangleq\left|\{\pi\in S_{q,\ell}\mid \exists x\in\Sigma^*,~x\vDash\pi \}\right|\mbox{, and}\\
R_{q,\ell}  &\triangleq \frac{\log_2 \cF_{q,\ell} }{\log_2(q^\ell!)}.
\end{align}
It is evident that~$\cF_{q,1}=q!$ for any~$q$. In addition, $\cF_{2,\ell}=0$ for every~$\ell\ge 2$, since (assuming $\Sigma=\{0,1\}$) the flow-conservation constraint $p_x(01^{\ell-1})+p_x(1^{\ell})=p_x(1^{\ell-1}0)+p_x(1^{\ell})$ implies $p_x(01^{\ell-1})=p_x(1^{\ell-1}0)$ contradicting the requirement that profile-vector entries be distinct. Therefore, the simplest set of parameters, that will be prominent in the sequel, is~$\ell=2$ and~$q=3$.

\section{Upper Bound}\label{section:upperBound}

For a given permutation~$\pi\in S_{q,\ell}$, a necessary condition for~$\pi$ to be feasible is given, and later used to obtain a bound on the number of feasible permutations. To formulate this condition, color an edge~$(a,b)$ of~$G_q^{\ell}$ in green if~$\pi(a)<\pi(b)$, and in red if~$\pi(a)>\pi(b)$. In addition, for a \textit{non-constant} string\footnote{That is, a string~$v$ that contains at least two distinct symbols.}~$v\in\Sigma^{\ell-1}$, let~$G(v)$ be an induced subgraph of~$G_q^\ell$ on the set~$T(v)\triangleq  \{\sigma_i v\}_{i=1}^q\cup \{v\sigma_i\}_{i=1}^q $, whose edges are colored as in~$G_q^\ell$.
For the next lemma, recall that a perfect matching in a graph is a vertex-disjoint set of edges which covers the entire vertex set of the graph.

\begin{lemma}\label{lemma:NecessaryConditionSubgraph}
	If there exists $v\in\Sigma^{\ell-1}$ such that~$G(v)$ contains an all-red perfect matching or an all-green perfect matching, then~$\pi$ is infeasible.
\end{lemma}

\begin{IEEEproof}
	Assume that there exists a subgraph $G(v)$ with an all-red perfect matching $\{(\sigma_i v,v\sigma_{\kappa(i)})\}_{i=1}^q$ for some permutation~$\kappa$ on~$[q]\triangleq \{1,\ldots ,q\}$. If there exists a string $x$ over $\Sigma$ which satisfies $\pi$, then since $x$ is a closed string, the profile vector~$p$ of~$x$ satisfies that $\sum_{i=1}^q p(\sigma_i v)=\sum_{i=1}^{q}p(v\sigma_i )$. However, since $p(\sigma_iv) > p(v\sigma_{\kappa(i)})$, we have that $\sum_{i=1}^q p(\sigma_i v)>\sum_{i=1}^{q}p(v\sigma_i)$, a contradiction. If $G(v)$ contains an all-green matching, the proof is similar.
\end{IEEEproof}

Let $\pi\vert_{T(v)}$ be the result of \textit{deleting} from~$\pi$ any element not in~$T(v)$. For example, if we take $q=3$, $\ell=2$, and~$\pi=(\texttt{AA},\texttt{AC},\texttt{CA},\texttt{GA},\texttt{AG},\texttt{CC},\texttt{CG},\texttt{GC},\texttt{GG})$, 
then $\pi\vert_{T(\texttt{A})}=(\texttt{AA},\texttt{AC},\texttt{CA},\texttt{GA},\texttt{AG})$. The given bound is derived by counting the number of \textit{infeasible} permutations on mutually disjoint sets $\{T(u_i)\}_{i=1}^k$, and estimating the number of permutations in~$S_{q,\ell}$ which contain an infeasible permutation on at least one~$T(u_i)$. To this end, the following lemmas are given.

\begin{lemma}\label{lemma:independentDisjoint}
	If $\{u_i\}_{i=1}^k\subseteq \Sigma^{\ell-1}$ is an independent set of vertices in~$G_q^{\ell-1}$, then the sets $T(u_i)$ are mutually disjoint.
\end{lemma}
\begin{IEEEproof}
	If there exist~$i$ and~$j$ in~$[k]$ such that $T(u_i)\cap T(u_j)\ne\varnothing$, then since $u_i\ne u_j$ it follows that there exist~$\sigma$ and~$\tau$ in~$\Sigma$ such that either $\sigma u_i= u_j\tau$ or $u_i\sigma = \tau u_j$. Without loss of generality assume that $\sigma u_i= u_j\tau$ and notice that in~$G_q^{\ell-1}$, $\sigma u_i$ is an edge entering node~$u_i$, and $u_j\tau$ is a edge leaving node $u_j$. Thus, $u_i$ and $u_j$ are connected by an edge, a contradiction.
\end{IEEEproof}

\begin{lemma}\label{lemma:allSubPermutations}
	Let $\{u_i\}_{i=1}^k\subseteq\Sigma^{\ell-1}$ be an independent set of non-constant vertices in~$G_q^{\ell-1}$. For any set of permutations $\{\pi_i\}_{i=1}^k$, where $\pi_i$ is a permutation on~$T(u_i)$, there are $\frac{q^\ell!}{(2q)!^k}$ permutations $\pi$ on~$\Sigma^{\ell}$ such that $\pi\vert_{T(u_i)}=\pi_i$ for all~$i\in[k]$.
\end{lemma}

\begin{IEEEproof}
	According to Lemma~\ref{lemma:independentDisjoint}, the sets $T(u_i)$ are mutually disjoint. Hence, the elements of $\cup_{i=1}^kT(u_i)$ may be arbitrarily interleaved such that for all~$i$, the relative order~$\pi_i$ of $T(u_i)$ is maintained. It is readily verified that the number of ways to interleave the elements of~$\{T(u_i)\}_{i=1}^k$, while maintaining the relative orders~$\{\pi_i\}_{i=1}^k$, equals the number of multi-permutations on the multi-set \[\{1,\ldots,1,2,\ldots,2,\ldots,k,\ldots,k\},\]where each element appears exactly~$2q$ times. Since the number of multi-permutations on this multi-set is given by the multinomial coefficient $(2q,\ldots,2q)!\triangleq \frac{(2qk)!}{(2q)!^k}$, and since the remaining~$q^\ell-k\cdot 2q$ elements may by inserted consecutively and arbitrarily into the resulting permutation, we have that the number of permutations~$\pi\in S_{q,\ell}$ such that~$\pi\vert_{T(u_i)}=\pi_i$ is
	\[
	\frac{(2qk)!}{(2q)!^k}\cdot\prod_{i=2qk+1}^{q^\ell} i=\frac{q^\ell!}{(2q)!^k}
	\]
\end{IEEEproof}

For any non-constant~$v\in\Sigma^{\ell-1}$, we now show that the set of monochromatic perfect matchings in~$G(v)$ is in one-to-one correspondence with the set~$\cS_q\subseteq\{ \pm 1 \}^{2q}$, where~$s\in\cS_q$ if and only if $\sum_{i=1}^{2q}s_i=0$ and $\sum_{i=1}^{j}s_i \ge 0$ for all~$j\in[2q]$. Consequently, the number of monochromatic perfect matchings is given by the $q$-th Catalan number $C_q=\frac{1}{q+1}\cdot{2q\choose q}$. For $s\in \cS_q$, let~$s^+\triangleq\{i\in[2q]\mid s_i=1\}$, and~$s^-\triangleq\{i\in[2q]\mid s_i=-1\}$.

\begin{lemma}\label{lemma:phiOnS}
	For~$s\in\cS_q$ there exists a bijection $\phi:s^+\to s^-$ such that $\phi(t)>t$ for all~$t\in s^+$.
\end{lemma}

\begin{IEEEproof}
	The function~$\phi$ is defined in a recursive manner. This definition relies on the recursive structure of~$\cS_q$, which is as follows. First we observe that the first entry of $s$ must be $s_1=1$. Next, denote the index of the left-most~$(-1)$-entry of~$s$ by~$m$, and let~$s'$ be the vector which results from~$s$ by replacing both~$s_1=1$ and $s_m=-1$ by~$0$. For any $j\in[2q]$ we have that
	\begin{align*}
	\sum_{i=1}^{j}s'_i=
	\begin{cases}
	j-1 & 1\le j<m\\
	\sum_{i=1}^{j}s_i & m\le j
	\end{cases}.
	\end{align*}
	Hence, it is readily verified that by omitting the first and $m$-th entry of~$s$ we obtain a vector in~$\cS_{q-1}$. Therefore, the function~$\phi$ may be defined by setting $\phi(1)=m$, omitting the first and $m$-th entry, and applying the same rule recursively.
\end{IEEEproof}

\begin{lemma}\label{lemma:numOfBadArrangements}
	For~$v\in\Sigma^{\ell-1}$, the number of infeasible permutations on~$T(v)$ is at least~$\frac{2}{q+1}\cdot(2q)!$.
\end{lemma}

\begin{IEEEproof}
	We count the number of permutations on~$T(v)$ that induce a monochromatic matching in~$G(v)$. Clearly, the existence of a monochromatic matching is oblivious to the \textit{internal} permutation on each of the sets~$\{\sigma_iv\}_{i=1}^q$ and~$\{v\sigma_i\}_{i=1}^q$. Further, the color of a monochromatic matching, if it exists, is uniquely determined by the lowest-ranking element of~$T(v)$. That is, if the lowest-ranking element belongs to~$\{\sigma_i v\}_{i=1}^q$, the matching is green, and otherwise it is red. Therefore, the number of permutations on~$T(v)$ that induce a monochromatic matching is 
	\[
	(q!)^2\cdot 2\cdot \kappa,
	\]
	where~$\kappa$ is the number of ways to interleave two fixed permutations on~$\{\sigma_iv\}_{i=1}^q$ and on~$\{v\sigma_i\}_{i=1}^q$, such that the identity of the lowest-ranking element (i.e., if it belongs to $\{\sigma_iv\}_{i=1}^q$ or to~$\{v\sigma_i\}_{i=1}^q$) is determined, and such that the resulting permutation induces a monochromatic matching. In what follows we show that~$\kappa$ equals the~$q$-th Catalan number, by employing the aforementioned set~$\cS_q$.
	
	For a permutation~$\pi_1$ on~$\{\sigma_iv\}_{i=1}^q$ and a permutation~$\pi_2$ on~$\{v\sigma_i\}_{i=1}^q$, let $C(\pi_1,\pi_2)$ be the number of permutations~$\pi$ on~$T(v)$ such that reducing~$\pi$ to $\{\sigma_iv\}_{i=1}^q$ results in~$\pi_1$, reducing~$\pi$ to~$\{v\sigma_i\}_{i=1}^q$ results in~$\pi_2$, and there exists a monochromatic green matching in~$G(v)$. By using a bijection to~$\cS_q$, it is shown that~$|C(\pi_1,\pi_2)|=C_q$.
	
	Given a vector~$s\in\cS_q$, let~$f(s)$ be the permutation of~$T(v)$ which results from replacing the~$1$ entries of~$s$ by the elements of~$\{\sigma_i v\}_{i=1}^q$, sorted by~$\pi_1$, and replacing the~$-1$ entries by the elements of~$\{v\sigma_i \}_{i=1}^q$, sorted by~$\pi_2$. According to Lemma~\ref{lemma:phiOnS}, in the resulting permutation~$f(s)$ any element in~$\{\sigma_i v  \}_{i=1}^q$ has a unique element in~$\{v\sigma_i \}_{i=1}^q$ that is higher ranked than it, and hence, $f(s)$ induces a green matching.
	
	Conversely, let~$\pi$ be a permutation on~$T(v)$ that induces a green matching~$M$, and hence, the smallest ranking element belongs to~$\{\sigma_i v\}_{i=1}^k$. Let~$g(\pi)$ be the vector in~$\{\pm 1\}^{2q}$ that results from replacing all entries of~$\pi$ which contain an element from~$\{\sigma_iv\}_{i=1}^q$ by~$1$, and replacing all entries which contain an element from~$\{v\sigma_i\}_{i=1}^q$ by~$-1$ (this is well defined since~$v$ has no self loops, and thus~$\{\sigma_iv\}_{i=1}^q\cap \{v\sigma_i\}_{i=1}^q=\varnothing$). 
	
	Assume for contradiction that~$g(\pi)$ induces a negative partial sum~$\sum_{i=1}^js_i<0$ for some~$j\in[2q]$. Since $s_1=1$, we may choose a \textit{minimal} such~$j$, where $\sum_{i=1}^{j-1}s_i=0$, $\sum_{i=1}^{j}s_i=-1$, and~$s_j=-1$. Consider the edges of~$M$ as connecting between the respective indices of~$s$, and note that since the matching is green, all edges~$(i,j)\in M$ satisfy that $s_i=1$, $s_j=-1$, and $i<j$. If there is no edge in the set~$[j-1]\times \{j,\ldots,2q\}$, we have that~$M$ contains an edge~$(j,t)$ with $t>j$ and $s_j=-1$, a contradiction. Similarly, if there exists an edge of~$M$ in~$[j-1]\times \{j,\ldots,2q\}$, since the number of~$1$ and~$-1$ entries among $s_1,\ldots,s_{j-1}$ is equal, 	we have that there exists a~$(-1)$-entry $s_t$, for~$t\le j-1$, such that $(t,m)\in M$  for some $m>t$. Once again, a contradiction.
	
	Since the above mappings~$g$ and~$f$ are injective, we have that~$|C(\pi_1,\pi_2)|=|\cS_q|=C_q$~\cite[p.~116]{LinWil01}. Note that the proof of this equality is identical for any~$\pi_1$ and~$\pi_2$, and that the proof is symmetric if the induced matching is red. Note also that a permutation~$\pi$ on~$T(v)$ cannot induce both a red and a green matching, since this would imply that~$\sum_{\sigma\in\Sigma}\pi(\sigma v)<\sum_{\sigma\in\Sigma}\pi(v\sigma)$ and $\sum_{\sigma\in\Sigma}\pi(\sigma v)>\sum_{\sigma\in\Sigma}\pi(v\sigma)$, a contradiction. Therefore, we have that the number of permutations on~$T(v)$ that induce a monochromatic matching is at least
	\[
	(q!)^2\cdot 2\cdot C_q = \frac{2}{q+1}\cdot(2q)!.
	\]
\end{IEEEproof}

To state the main result of this section, we introduce the \textit{loopless independence number} $\alpha^*(q,\ell)$ of $G_q^\ell$~\cite{Lic06}, which is the size of the largest subset of nodes that contains no internal edges and no self-loops (i.e., the nodes are non-constant). A lower bound on~$\alpha^*$ may be easily derived from~\cite{Lic06}.

\begin{lemma} [\cite{Lic06}]\label{lemma:alphaStar}
	$\alpha^*(q,\ell)\ge \frac{q^\ell-q^{\ell-2}}{4}$, for all $q\geq 2$ and $\ell\geq 2$.
\end{lemma}

\begin{IEEEproof}
	According to~\cite[Proposition~5.2]{Lic06} we have that if~$q\ge 2$ and $\ell\ge 3$, then~$\alpha^*(q,\ell)\ge {q\cdot\alpha^*(q,\ell-1)}$. By applying this claim recursively, we have that $\alpha^*(q,\ell)\ge q^{\ell-2}\cdot\alpha^*(q,2)$.
	According to~\cite[Sec.~5]{Lic06} we have that if~$q\ge 2$ then $\alpha^*(q,2)\ge \frac{q^2-1}{4}$. The result follows from combining these claims.
\end{IEEEproof}

\begin{theorem}\label{theorem:bound}
	For all $q\geq 3$ and $\ell\geq 3$, the number of feasible permutations is at most
	\[
	\cF_{q,\ell}\leq q^{\ell}!\cdot\left(\frac{q-1}{q+1}\right)^{\frac{1}{4}(q^{\ell-1}-q^{\ell-3})}.
	\]
\end{theorem}

\begin{IEEEproof}
	Let~$\alpha^*=\alpha^*(q,\ell-1)$, and let~$\{u_i\}_{i=1}^{\alpha^*}\subseteq\Sigma^{\ell-1}$ be a maximum loopless independent set in~$G_q^{\ell-1}$. Since it is loopless, it follows that none of~$\{u_i\}_{i=1}^{\alpha^*}$ is constant. Hence, Lemma~\ref{lemma:allSubPermutations} implies that for every set~$\{\pi_i\}_{i=1}^{\alpha^*}$ of permutations, where~$\pi_i$ is a permutation on~$T(u_i)$, there are~$\frac{q^\ell!}{(2q)!^{\alpha^*}}$ permutations~$\pi\in S_{q,\ell}$, such that~$\pi\vert_{T(u_i)}=\pi_i$ for all~$i\in[\alpha^*]$.
	
	Since by Lemma~\ref{lemma:numOfBadArrangements} the number of infeasible permutations on each~$T(u_i)$ is at least~$\frac{2}{q+1}(2q)!$, it follows that there are at least 
	\begin{align*}
	(2q)!^{\alpha^*}-\left(\frac{q-1}{q+1} \cdot (2q)!\right)^{\alpha^*}=\left(1-\left(\frac{q-1}{q+1}\right)^{\alpha^*} \right)(2q)!^{\alpha^*}
	\end{align*}
	sets~$\{\pi_i\}_{i=1}^{\alpha^*}$ which contain at least one infeasible permutation. This implies, together with Lemma~\ref{lemma:alphaStar}, that the number of infeasible permutations~$\pi\in S_{q,\ell}$ is at least
	\begin{align*}
	q^\ell!-\cF_{q,\ell}\geq \left(1-\left(\frac{q-1}{q+1} \right)^{\alpha^*} \right)\cdot (2q)!^{\alpha^*}\cdot\frac{q^\ell!}{(2q)!^{\alpha^*}}=\left(1-\left(\frac{q-1}{q+1} \right)^{\alpha^*} \right)\cdot q^\ell!\ge \left(1-\left(\frac{q-1}{q+1}\right)^{\frac{q^{\ell-1}-q^{\ell-3}}{4}}\right)\cdot q^{\ell}!
	\end{align*}
\end{IEEEproof}

For~$\ell=2$, similar techniques do not hold since~$\alpha^*=\alpha(q,1)=0$. However, we may apply Lemma~\ref{lemma:numOfBadArrangements} directly without enhancing it by the loopless independence number of~$G_q^{\ell-1}$. This results in the following weaker bound, whose proof is similar to that of Theorem~\ref{theorem:bound}, and is therefore omitted.

\begin{theorem}
	For all~$q\ge 3$ we have~$\cF_{q,2}\le (q^2)!\cdot\frac{q-1}{q+1}$.
\end{theorem}

Theorem~\ref{theorem:bound} readily implies that if either~$q$ or~$\ell$ goes to infinity, the fraction of feasible permutations goes to zero. However, in terms of rate, both $R_{q,\ell}\xrightarrow[q\to\infty]{} 1$ and $R_{q,\ell}\xrightarrow[\ell\to\infty]{} 1$, leaving the possibility for high-rate schemes wide open. In the sequel, a set of feasible permutations is constructed, whose rate is bounded from below by a constant when~$\ell$ is fixed and~$q$ tends to infinity (Lemma~\ref{lemma:Rq2} in Section~\ref{section:encoding}).

\section{A Polynomial Algorithm for Deciding Feasibility}\label{section:polynomialAlgorithm}
In this section it is shown that a given permutation $\pi\in S_{q,\ell}$ can be decided to be feasible or not in time polynomial in~$q^\ell$ (i.e., polynomial in the length of the permutation). If it is feasible, the time complexity of producing a string which satisfies it depends on its minimal length, a topic which is studied in Section~\ref{section:length}. The given algorithm relies on the following simple claim.

\begin{lemma}\label{lemma:piChi}
	A permutation~$\pi\in S_{q,\ell}$ is feasible if and only if there exists a nonnegative vector~$\chi\in\bR^{q^\ell}$ such that $\pi\vDash\chi$ and for all~$v\in\Sigma^{\ell-1}$, $\sum_{\sigma\in\Sigma}\chi(v\sigma)=\sum_{\sigma\in\Sigma}\chi(\sigma v)$.
\end{lemma}

\begin{IEEEproof}
	If~$\pi$ is feasible, then by the definition of feasibility, there exists $x\in\Sigma^*$ such that its~$\ell$-profile vector~$p_x$ satisfies~$\pi$, and hence, setting~$\chi=p_x$ suffices.
	Conversely, given a non-negative vector~$\chi$ that satisfies $\pi$, as well as flow conservation, we make the following observation: for any $\alpha,\beta\in\bR$, $\alpha,\beta\geq 0$, we have that $\alpha\chi+\beta\mathbf{1}$ is a non-negative vector that satisfies $\pi$ as well as flow conservation. Choose $\alpha$ and $\beta$ such that $\chi'\triangleq \alpha\chi+\beta\mathbf{1}$ has $\chi(v)\geq 1$ and $|\chi'(v)-\chi'(v')|\geq 3$ for all $v,v'\in\Sigma^\ell$, $v\neq v'$.
	
    By \cite[Appendix]{EliMeySch16}, there exists an \emph{integer} flow-conserving vector $\chi''$ where $\lfloor\chi'(v)\rfloor\leq \chi''(v)\leq \lceil\chi'(v)\rceil +1$. By our choice of $\alpha$ and $\beta$, all the entries of $\chi''$ are positive, distinct, and retain their rank, i.e., $\chi''$ satisfies $\pi$. By Section \ref{section:preliminaries}, this suffices for the existence of a string $x\in\Sigma^*$ whose profile vector is $p_x=\chi''$, hence $x$ satisfies $\pi$.
\end{IEEEproof}

Given a permutation~$\pi$, Lemma~\ref{lemma:piChi} gives rise to the following linear programming algorithm for deciding its feasibility.

\begin{itemize}
	\item \textbf{Variables:}~$\{\chi(v) \mid v\in\Sigma^{\ell}\}$.
	\item \textbf{Objective:} None.
	\item \textbf{Constriants:}
	\begin{itemize}
		\item For all~$v\in\Sigma^\ell$, $\chi(v)\ge 1$.
		\item For all distinct~$u$ and~$v$ in $\Sigma^\ell$ such that $\pi(u)>\pi(v)$, $\chi(u) \geq \chi(v)+1$.
		\item For all~$v\in\Sigma^{\ell-1}$, $\sum_{\sigma\in\Sigma}\chi(v\sigma)=\sum_{\sigma\in\Sigma}\chi(\sigma v)$.
	\end{itemize}
\end{itemize}
According to Lemma~\ref{lemma:piChi}, determining the feasibility of this system is equivalent to determining the feasibility of the given permutation~$\pi$.

Note that the number of variables is~$q^\ell$, and the number of constraints is~$q^\ell+{q^\ell\choose 2}+q^{\ell-1}$, which is polynomial in~$q^\ell$. Hence, the feasibility question may be solved in polynomial time (in~$q^\ell$).

Since the coefficients in the linear program are all rational, the feasible solutions contain a solution~$\chi$ that is rational in all of its entries. One may then find a corresponding string by the techniques that are mentioned in Section~\ref{section:preliminaries}.

\section{Encoding Algorithms}\label{section:encoding}
In this section encoding algorithms are given for any~$\ell$ and any~$q$. These algorithms provide a lower bound on the number of feasible permutations for the respective parameters. Since an additive structure of the alphabet is required, it is assumed in this section that~$\Sigma=\bZ_q$, the set of integers modulo~$q$.

According to Lemma~\ref{lemma:piChi} and its subsequent discussion, to come up with a feasible permutation it suffices to provide the corresponding vector~$\chi$. From this vector the corresponding permutation and a suitable string may be computed by either the randomized or the deterministic algorithms that are mentioned after Lemma~\ref{lemma:piChi}. Hence, the algorithms that are detailed in this section focus on providing the vector~$\chi$.

To clearly describe the constraints on the vector~$\chi$, by abuse of notation it will be considered either as a vector in~$\bR^{q^\ell}$ or as a \emph{matrix} in~$\bR^{q^{\ell-1}\times q^{\ell-1}}$. That is, for a given string~$u\in \bZ_q^\ell$, the notation~$\chi(u)$ stands for the~$u$-th entry of~$\chi$ when seen as a vector, and for two strings $w,w'\in\Sigma^{\ell-1}$, the notation~$\chi(w,w')$ stands for the~$(w,w')$-entry when seen as a matrix, where
\begin{align}\label{equation:chiMatrix}
\chi(w,w')\triangleq
\begin{cases}
\chi(u) & \text{if $(w,w')$ is a $u$-labeled edge in $E(G_q^{\ell-1})$,}\\
0 & \text{else.}
\end{cases}
\end{align}
Using this notation, for~$v\in\bZ_q^{\ell-1}$ the constraint $\sum_{\sigma\in\bZ_q}\chi(v\sigma)=\sum_{\sigma\in\bZ_q}\chi(\sigma v)$ may be written as $\sum_{u\in\bZ_q^{\ell-1}}\chi (v,u)=\sum_{u\in\bZ_q^{\ell-1}}\chi (u,v)$, i.e., the $v$-th row sum equals the~$v$-th column sum. A vector (matrix)~$\chi$ which satisfies these constraints and satisfies a permutation, is called a \emph{feasible vector (matrix)}. Note that if a feasible vector~$\chi$ has no zero entries, then the support of the corresponding matrix is identical to the support of the adjacency matrix of~$G_q^{\ell-1}$.

We present two algorithms in this section. The first fixes a window size of $\ell=2$, and recursively increases $q$. The second algorithm builds on the first one, and extends the window size $\ell$.

\subsection{A recursive encoding algorithm for~$\ell=2$ and any~$q$}\label{section:ell=2}
This algorithm operates recursively on~$q$, where the base case is~$q=3$. To resolve the base case, a repository of all feasible permutations (or their corresponding feasible matrices) for~$q=3$ and~$\ell=2$ should be maintained. Using a computer program which is based on Lemma~\ref{lemma:piChi}, this repository was constructed within a few minutes on a laptop computer, and its size was discovered to be~$f_{3,2}\triangleq 30240$. For future use, it is convenient to assume that all matrices in this repository contain strictly positive integer entries, which is always possible in light of the proof of Lemma~\ref{lemma:piChi} and the fact that $G^1_q$ is the complete graph on $q$ nodes. In Algorithm~\ref{algorithm:A_q} which follows, the information vector is taken from
\begin{align}\label{equation:I_q}
I_q\triangleq [f_{3,2}]\times (S_4\times K_4)\times\cdots\times (S_q\times K_q),
\end{align}
where for any positive integer~$m$, the set~$K_m$ consists of all~$(m^2-m+1)$-bit vectors with Hamming weight~$m$. Given~$t\in K_m$, a vector~$x$ of length~$(m-1)^2$ and a vector~$y$ of length~$m$, let~$t(x,y)$ be the vector which results from replacing the~$1$-entries in~$t$ by the entries of~$x$ and the~$0$-entries by~$y$, while maintaining their original order.

\begin{algorithm}[H]
	\SetKwInOut{Input}{Input}
	\SetKwInOut{Output}{Output}
	\KwData{A repositpry~$R$ of~$f_{3,2}$ integer feasible matrices for all feasible permutations in~$S_{3,2}$.}
	\Input{An information vector~$v\in I_q$.}
	\Output{A feasible matrix $\chi\in\bN^{q\times q}$ (which represents a feasible vector~$\chi\in\bN^{q^2}$).}
	\lIf{$q=3$}{return the~$v$-th feasible matrix in~$R$}\label{algorithm1Line:base}
	Denote~$v=(v',(\pi_q,t_q))$ for $v'\in I_{q-1}$, $\pi_q \in S_{q}$, and~$t_q\in K_q$.\\
	Apply $A_{q-1}(v')$ to get a matrix~$\chi'$, and for~$i,j\in\bZ_{q-1}$ denote $x_{i,j}\triangleq (q+1)\cdot(\chi')_{i,j}$.\label{algorithm1Line:multiply}\\
		Choose~$y\triangleq(y_0,\ldots,y_{q-1})\in\bN^{q}$ that is ordered by~$\pi_q$, such that $\sum_{i=0}^{q-1}y_i$ is minimal, and such that~$t_q(\mbox{sort}(y),\mbox{sort}((q+1)\cdot\chi')$ is sorted\footnotemark.\label{algorithm1Line:Choose}\\
		For~$\varepsilon\triangleq\frac{1}{q}$, let
		\begin{align*}
			\chi\triangleq
			\begin{pmatrix}
			 x_{0,0}                                  & x_{0,1} & \cdots & x_{0,q-2} & y_{0}\\
			 x_{1,0}+\varepsilon					    & x_{1,1} & \cdots & x_{1,q-2}&y_1-\varepsilon\\
			 x_{2,0}+\varepsilon					    & x_{2,1} & \cdots & x_{2,q-2}&y_2-\varepsilon\\
			 \vdots  								    & \vdots  & \ddots & \vdots & \vdots\\
			 x_{q-2,0}+\varepsilon			    & x_{q-2,1} & \cdots & x_{q-2,q-2}& y_{q-2}-\varepsilon\\
			 y_{0}-(q-2)\varepsilon & y_1 & \cdots & y_{q-2} &  y_{q-1}
			\end{pmatrix}.
		\end{align*}\\
		\Return~$q\cdot\chi$.\label{algorithm1Line:return}
	\caption{$A_q(v)$, an encoding algorithm for~$\ell=2$ and any~$q$.}
	\label{algorithm:A_q}
\end{algorithm}
\footnotetext{For a real vector~$v$, let~$\mbox{sort}(v)$ be the output of applying a sorting algorithm to~$v$.}

In a nutshell, each recursive step of Algorithm~\ref{algorithm:A_q} assigns values for entries in~$\chi$ that correspond to strings which contain the newly added symbol of the alphabet. These new strings are ordered according to~$\pi_q$, which is taken from the information vector, and are interleaved with the entries of~$\chi'$ according to~$t_q$. Following the next example, the correctness of Algorithm~\ref{algorithm:A_q} is proved in detail.

\begin{example}\label{example:Alg1}
	Assume that for~$q=4$, the given information vector is~$v=(v',(\pi_4,t_4))$ where
	\begin{align*}
		\pi_4&=(2,3,4,1),\\
		t_4  &=(0,0,1,1,1,0,0,1,0,0,0,0,0),
	\end{align*}
	and~$v'$ is the index in~$[f_{3,2}]$ of the permutation~$(00,01,10,20,02,11,12,21,22)$ (which results from~\eqref{equation:examplePi} by substituting~$0$ for~$\texttt{A}$, $1$ for~$\texttt{C}$, and $2$ for~$\texttt{G}$). Assume that by applying~$A_3(v')$, we obtain the feasible matrix
	\begin{align}\label{equation:chi'}
		\chi'=
		\begin{pmatrix}
		1 & 2 & 5\\
		3 & 6 & 7\\
		4 & 8 & 9
		\end{pmatrix},
	\end{align}
	where the rows and columns are identified lexicographically by~$\bZ_3$. Clearly, choosing~$y=(12,13,21,11)$ suffices since it is ordered by~$\pi_4$, since
	\begin{align*}
		t_4(\mbox{sort}(y),\mbox{sort}(5\cdot\chi'))=(5,10,11,12,13,15,20,21,25,30,35,40,45),
	\end{align*}
	and since~$\sum_{i=0}^{3}y_i$ is clearly minimal. Thus, it follows that
	\begin{align*}
		\chi=
		\begin{pmatrix}
			5 & 10 & 25 & 12 \\
			15+\frac{1}{4} & 30 & 35 & 13-\frac{1}{4}\\
			20+\frac{1}{4} & 40 & 45 & 21-\frac{1}{4}\\
			12-2\cdot \frac{1}{4}& 13 & 21 & 11
		\end{pmatrix},
	\end{align*}
	and hence the output matrix is
	\begin{align*}
		4\cdot\chi=
		\begin{pmatrix}
			20 & 40 & 100 & 48\\
			61 & 120 & 140 & 51\\
			81 & 160 & 180 & 83\\
			46 & 52 & 84 & 44
		\end{pmatrix},
	\end{align*}
	which is an integer feasible matrix for 
	\begin{align*}
		(00,01,33,30,03,13,31,10,20,23,32,02,11,12,21,22).
	\end{align*}
\end{example}

First, notice that by the choice of~$\varepsilon$ and by the assumption on~$R$, the matrix which is returned in either Line~\ref{algorithm1Line:base} or Line~\ref{algorithm1Line:return} has positive integer entries. Hence, the multiplication of~$\chi'$ by~$q+1$ in Line~\ref{algorithm1Line:multiply} provides~$q$ distinct integers between every two entries of~$\chi'$. In turn, this enables the choice of~$y_0,\ldots,y_{q-1}$, and their interleaving with the entries of~$\chi'$, in any possible way.

Second, to verify that the entries of~$\chi$ are distinct, recall that $\{y_0,\ldots,y_{q-1}\}\cup\{x_{i,j}\vert i,j\in\bZ_{q-1} \}$ is a set of distinct integers. Hence, since~$\frac{1}{q}\le\frac{1}{3}$ and~$1-(q-2)\varepsilon>\varepsilon$, it follows that the entries of~$\chi$ are distinct. To prove the correctness of the algorithm, it ought to be shown that the row and column sum constraint from Lemma~\ref{lemma:piChi} is satisfied, and that different information vectors result in different permutations.

\begin{lemma}\label{lemma:l=2AlgoCorrectness1}
  Let $q\cdot\chi$ be the output of Algorithm~\ref{algorithm:A_q}. Then for all~$\tau\in\bZ_q$, $\sum_{\sigma\in\bZ_q}\chi(\tau,\sigma)=\sum_{\sigma\in\bZ_q}\chi(\sigma,\tau)$.
\end{lemma}

\begin{IEEEproof}
	First, it is evident that
	\begin{align*}
	\mbox{sum of column }q-1 &= \sum_{i\in\bZ_q}y_i -\sum_{i=1}^{q-2}\varepsilon=\sum_{i\in\bZ_q}y_i -(q-2)\varepsilon=\mbox{sum of row }q-1,
	\end{align*}
	and since~$\chi'$ it a feasible matrix, it follows that
	\begin{align*}
	\mbox{sum of column~0} &= y_0-(q-2)\varepsilon+x_{0,0}+\sum_{i=1}^{q-2}(x_{i,0}+\varepsilon) =y_0+\sum_{i=0}^{q-2}x_{0,i} =\mbox{sum of row~0}.
	\end{align*}
	Further, for~$1\le i\le q-2$,
	\begin{align*}
	\mbox{sum of row }i &=( x_{i,0}+\varepsilon)+\sum_{j=1}^{q-2}x_{i,j}+(y_i-\varepsilon)= \sum_{j=0}^{q-2}x_{j,i}+y_{i} =\mbox{sum of column }i.
	\end{align*}
\end{IEEEproof}

\begin{lemma}\label{lemma:A_qDistinct}
	If~$u$ and~$v$ are distinct information vectors in~$I_q$ such that $A_q(u)=\chi_u,A_q(v)=\chi_v$ and $\chi_u\vDash \pi_u,\chi_v\vDash \pi_v$ for some permutations~$\pi_u$ and~$\pi_v$, then $\pi_u\ne \pi_v$.
\end{lemma}

\begin{IEEEproof}
	Note that according to the choice of~$\varepsilon$, the relative order among the entries of~$\chi'$ is preserved. Hence, if~$u_i\ne v_i$ for some entry~$i\ge 2$, then either the respective permutations between the $y_j$-s in stage~$i$ of the algorithm are distinct, or the interleaving of the~$y_j$-s in the entries of~$\chi'$ are distinct. In either case, it follows that there exists a pair of corresponding strings (either both new, or one is new and the other is old) on whom~$\pi_v$ and~$\pi_u$ disagree; and since this disagreement persists throughout the entire algorithm, the result follows. If~$u$ and~$v$ disagree only on their first (leftmost) entry, the proof is similar.
\end{IEEEproof}

\begin{corollary}\label{corollary:numFeasible_ell=2}
	For any~$q\ge 3$,
	\begin{align*}
		\cF_{q,2}\ge f_{3,2}\cdot\prod_{j=4}^{q}\left(j!\cdot {j^2-j+1\choose j} \right).
\end{align*}
\end{corollary}

Even though taking~$q$ to infinity is artificial when DNA storage is discussed, it is inevitable if one wishes to estimate the contribution of Algorithm~\ref{algorithm:A_q}. Hence, the following lemma is given, where the proof is deferred to Appendix~\ref{section:omittedProofs}.

\begin{lemma}\label{lemma:Rq2}
	$\lim_{q\to\infty}R_{q,2}\ge\frac{1}{2}$.
\end{lemma}

\subsection{A recursive encoding algorithm for any~$\ell$ and any~$q$}\label{section:ellAtleast3}
In this section, the recursive algorithm from Subsection~\ref{section:ell=2} is used to obtain an encoding algorithm for any~$\ell$ and any~$q$. Inspired by~\cite{Lem70} and~\cite{Liu90}, the recursion at stage~$\ell$ relies on embedding the feasible matrix from stage~$\ell-1$ in \textit{homomoprhic pre-images} of~$G_q^{\ell-1}$ in~$G_q^\ell$, and breaking ties that emerge according to the information vector.

\begin{definition}\cite{GroYel04}
	For graphs~$G$ and~$H$, a function $f:V(G)\to V(H)$ is a (graph) homomorphism if for each pair of vertices~$u,v$ in~$V(G)$, if $(u,v)$ is an edge in~$G$ then~$(f(u),f(v))$ is an edge in~$H$.
\end{definition}

The algorithm which follows relies on the following homomorphism, and yet, many other homomorphisms exist, and either of them may be used similarly.
\begin{align*}
&D_\ell:\bZ_q^\ell\to\bZ_q^{\ell-1}\\
&D_\ell(v)\triangleq (v_0+v_1,v_1+v_2,\ldots,v_{\ell-2}+v_{\ell-1})\mbox{, and}\\
&D:\bZ_q^*\to\bZ_q^*\\
&D(v)\triangleq D_{|v|}(v).
\end{align*}
It is an easy exercise to prove that for any~$\ell$, the function~$D$ is a~$q$ to~$1$ surjective homomorphism from~$G_q^\ell$ to~$G_q^{\ell-1}$. That is, for every~$u\in \bZ_q^{\ell-1}$, the set~$D^{-1}(u)$ contains exactly~$q$ elements. Moreover, it is readily verified that this set may be written as 
\begin{align}\label{equation:Dinverse}
D^{-1}(u)=\{v_0,\ldots,v_{q-1} \}\mbox{, such that }v_{i,0}=i\mbox{ for all }i\in\bZ_q.
\end{align}
Similarly, every~$u\in\bZ_q^{\ell-1}$ satisfies that
\begin{align}\label{equation:flipSigma}
\nonumber\{D(\sigma u) \}_{\sigma\in\bZ_q}&=\{\sigma D(u)\}_{\sigma\in\bZ_q}\mbox{, and}\\
\{D(u\sigma) \}_{\sigma\in\bZ_q}&=\{ D(u)\sigma\}_{\sigma\in\bZ_q}.
\end{align}
The information vector is taken from the set~$J_q^\ell$, which is defined as follows.
\begin{align}\label{equation:DefJ_qell}
\nonumber J_q^\ell&\triangleq I_q\times \cP_3\times \cP_{4}\times \ldots \times \cP_\ell\mbox{, where}\\
\nonumber \cP_i&\triangleq\{P\mid P:\cA_{i}\to S_{\bZ_q} \}\mbox{, and}\\
\cA_i &\triangleq \{u\in\bZ_q^{i-1}\mid 0\notin \{u_0,u_{i-2} \}\} \mbox{ for all } i\in\{3,\ldots,\ell \}.
\end{align}
That is, an information vector~$v\in J_q^\ell$ is of the form $v=(v',P_{3},\ldots,P_{\ell})$, where~$P_i$ is a function from~$\cA_i$ to~$S_{\bZ_q}$, and $v'\in I_q$ (see Subsection~\ref{section:ell=2}). In addition, for~$i$ and~$j$ in~$\bZ_q$, let~$\qnum{i}{j}_q \triangleq i+j\cdot q+1$.

In stage~$\ell$, Algorithm~\ref{algorithm:B_q^ell} relies on embedding the matrix~$T$, which results from stage~$\ell-1$, in entries of~$\chi_\ell$ that correspond to homomorphic pre-images of~$G_q^{\ell-1}$ in~$G_q^\ell$. Since the homomorphism~$D$ is~$q$ to~$1$, this results in~$q^{\ell-1}$ sets of entries in~$\chi_\ell$, of~$q$ elements each, that contain identical entries. These equalities are broken by adding small constants to each set, where these constants are ordered according to the permutations in~$S_q$ that appear in the current entry of the information vector. To maintain the row and column sum constraint (Lemma~\ref{lemma:piChi}), the additions will be excluded from a single entry in each row and each column, where in turn, these excluded entries will be adjusted to cancel out the additions in their respective row or column.

\begin{algorithm}[H]
	\SetKwInOut{Input}{Input}
	\SetKwInOut{Output}{Output}
	\Input{An information vector~$v\in J_q^\ell$.}
	\Output{A feasible matrix $\chi\in\bN^{q^{\ell-1}\times q^{\ell-1}}$ (which represents a feasible vector~$\chi\in\bN^{q^\ell}$).}
	\lIf{$\ell=2$}{return~$A_q(v)$}
	Denote~$v=(v',P)$, where $P\in\cP_\ell$ and $v'\in J_{q}^{\ell-1}$.\\
	Set $T=2\cdot B^{\ell-1}_q(v')$.\label{algorithm2Line:multiply}\\
	\ForEach{$v\in\bZ_q^\ell$}{
		Denote~$v=v_0v'$, where~$v_0\in\bZ_q$ and~$v'\in\bZ_q^{\ell-1}$.\\
		Denote $D(v)\triangleq w=(w_0,w',w_{\ell-2})$, where~$w_0,w_{\ell-2}\in\bZ_q$ and~$w'\in\bZ_q^{\ell-3}$.\\
		Define 
		\[
		\chi_\ell(v)=
		\begin{cases}
			T(D(v))+\frac{P(D(v))(v_0)}{q^{\qnum{w_0}{w_{\ell-2}}}} & \mbox{(1)~if }D(v)\in \cA_\ell\\
			T(D(v))-\sum_{\mu\ne 0}\frac{P(\mu,w',w_{\ell-2})(\mu+v_0)}{q^{\qnum{\mu}{w_{\ell-2}}}} & \mbox{(2)~if }w_0=0,w_{\ell-2}\ne 0\\
			T(D(v))-\sum_{\tau\ne 0}\frac{P(w_0,w',\tau)(v_0)}{q^{\qnum{w_0}{\tau}}} & \mbox{(3)~if }w_0\ne 0,w_{\ell-2}=0\\
			T(D(v))+\sum_{\mu\ne 0}\sum_{\tau\ne 0}\frac{P(\mu,w',\tau)(\mu+v_0)}{q^{\qnum{\mu}{\tau}}} & \mbox{(4)~if }w_0=w_{\ell-2}=0 \\
		\end{cases}
		\]\label{algorithm2Line:define}
	}
	\Return~$q^{q^2}\cdot \chi_\ell$.\label{algorithm2Line:return}
	\caption{$B_q^\ell(v)$, an encoding algorithm for any~$\ell$ and any~$q$.}
	\label{algorithm:B_q^ell}
\end{algorithm}

\begin{example}\label{example:Alg2}
	For~$q=3$ and~$\ell=3$, notice that~$\cA_3=\{ (11),(12),(21),(22) \}$ and that~$\cP_3=\{ P\vert P:\cA_3\to S_{\bZ_3} \}$. 
	The output matrix in this case is $\chi=q^{q^2}\chi_3$, where~$\chi_3$ is defined by using~$T=2B_3^2(v')$. The three leftmost columns of~$\chi_3$ are as follows.
	
	\begin{center}
		\begin{tabular}{|c|c|c|c|c|c|c|c|c|c|} \cline{2-4}
		\multicolumn{1}{c|}{}  & 00 & 01 & 02 \\ \hline
		00 &  $T(00)+\sum_{\mu\ne 0}\sum_{\tau\ne 0}\frac{P(\mu\tau)(\mu)}{q^{\qnum{\mu}{\tau}}}$ &  $T(01)-\sum_{\mu\ne 0}\frac{P(\mu1)(\mu)}{q^{\qnum{\mu}{1}}}$ & $T(02)-\sum_{\mu\ne 0}\frac{P(\mu2)(\mu)}{q^{\qnum{\mu}{2}}}$  \\ \hline
		01 &  - &  - &  -  \\ \hline
		02 &  - &  - &  -  \\ \hline
		10 &  $T(10)-\sum_{\tau\ne 0}\frac{P(1\tau)(1)}{q^{\qnum{1}{\tau}}}$ & $T(11)+\frac{P(11)(1)}{q^{\qnum{1}{1}}}$ &  $T(12)+\frac{P(12)(1)}{q^{\qnum{1}{2}}}$  \\ \hline
		11 &  - &  - &  -  \\ \hline
		12 &  - &  - &  -  \\ \hline
		20 &  $T(20)-\sum_{\tau\ne 0}\frac{P(2\tau)(2)}{q^{\qnum{2}{\tau}}}$ &  $T(21)+\frac{P(21)(2)}{q^{\qnum{2}{1}}}$ &  $T(22)+\frac{P(22)(2)}{q^{\qnum{2}{2}}}$  \\ \hline
		21 &  - &  - &  -  \\ \hline
		22 &  - &  - &  -  \\ \hline
		\end{tabular}
	\end{center}
Notice that for each nonzero entry in the above table, the entries of~$\chi_3$ that share the same row or column and are \textit{not} listed above are zero. Hence, one can verify that the row and column sums coincide with those of~$\tau$. To witness the distinctness of entries in a homomorphic preimage of~$D$, consider for example~$D^{-1}(11)=\{ 010,101,222\}$, and notice that
\begin{align*}
	\chi_3(010) &= T(11)+\frac{P(11)(0)}{q^{q+2}}\\
	\chi_3(101) &= T(11)+\frac{P(11)(1)}{q^{q+2}}\\
	\chi_3(222) &= T(11)+\frac{P(11)(2)}{q^{q+2}},
\end{align*}
which are clearly distinct since~$P(11)$ is a permutation. Yet another example is given by considering~$D^{-1}(20)=\{ 021,112,200\}$, for which 
\begin{align*}
\chi_3(021) &= T(20)-\sum_{\tau\ne 0}\frac{P(2\tau)(0)}{q^{\qnum{2}{\tau}}}=T(20)-\frac{P(21)(0)}{q^{q+3}}-\frac{P(22)(0)}{q^{2q+3}}\\
\chi_3(112) &= T(20)-\sum_{\tau\ne 0}\frac{P(2\tau)(1)}{q^{\qnum{2}{\tau}}}=T(20)-\frac{P(21)(1)}{q^{q+3}}-\frac{P(22)(1)}{q^{2q+3}}\\
\chi_3(200) &= T(20)-\sum_{\tau\ne 0}\frac{P(2\tau)(2)}{q^{\qnum{2}{\tau}}}=T(20)-\frac{P(21)(2)}{q^{q+3}}-\frac{P(22)(2)}{q^{2q+3}}.
\end{align*}
Similarly, the above entries are distinct, e.g., on the~$(q+3)$-rd digit of the fractional~$q$-ary expansion, since~$P(21)$ is a permutation.
\end{example}

To show the correctness of Algorithm~\ref{algorithm:B_q^ell}, it suffices to show that~$\chi_\ell$ complies with the row and column sum constraint of Lemma~\ref{lemma:piChi}, and that its entries are distinct.

\begin{lemma}\label{lemma:B_q^ellConstraints}
	In Algorithm~\ref{algorithm:B_q^ell} we have that $\sum_{\sigma\in\bZ_q}\chi_\ell(\sigma u)=\sum_{\sigma\in\bZ_q}\chi_\ell(u\sigma)$ for every~$u\in\bZ_q^{\ell-1}$ and every integer~$\ell\ge 3$.
\end{lemma}

\begin{IEEEproof}
	The correctness in stage~$\ell$ follows from the correctness in stage~$\ell-1$, and the following four technical facts. These facts show that the additions to the entries of~$T$ cancel out, and hence the row and column sums of~$\chi_\ell$ are equal to those of~$T$, which satisfies the row and column sum constraint. First, note that any~$u\in\bZ_q^{\ell-1}$ satisfies exactly one of the following two cases.
	\begin{itemize}
		\item [Case~A1.] $D(u)_{\ell-3}\ne 0$. For any~$\sigma\in\Sigma$, the string~$v= \sigma u$ satisfies that~$w_{\ell-2}\ne 0$ and that~$w_0=0$ if and only if~$\sigma=-u_1$. Therefore,~$\chi_{\ell}(v)$ is computed by either~$(1)$ or~$(2)$. Notice that
			\begin{align*}
				\sum_{\sigma\in\bZ_q}\chi_\ell(\sigma u)&=\sum_{\sigma\ne -u_1}\left(T(D(\sigma u))+\frac{P(D(\sigma u))(\sigma)}{q^{\qnum{\sigma+u_1}{u_{\ell-3}+u_{\ell-2}}}} \right)\\
				&\phantom{=}+T\left(D(-u_1,u) \right)-\sum_{\mu\ne 0}\frac{P(\mu,D(u))(\mu+(-u_1))}{q^{\qnum{\mu}{u_{\ell-3}+u_{\ell-2}}}}\\
				&=\sum_{\sigma\ne -u_1}\left(T(D(\sigma u))+\frac{P(\sigma+u_1,D(u))(\sigma)}{q^{\qnum{\sigma+u_1}{u_{\ell-3}+u_{\ell-2}}}} \right)\\
				&\phantom{=}+T\left(D(-u_1,u) \right)-\sum_{\mu\ne -u_1}\frac{P(\mu+u_1,D(u))(\mu)}{q^{\qnum{\mu+u_1}{u_{\ell-3}+u_{\ell-2}}}}\\
				&=\sum_{\sigma\in\bZ_q}T(D(\sigma u))\overset{\eqref{equation:flipSigma}}{=}\sum_{\sigma\in\bZ_q}T(\sigma D( u))=\sum_{\sigma\in\bZ_q}T( D( u)\sigma).
			\end{align*}
			
		\item[Case~A2.] $D(u)_{\ell-3}=0$. For any~$\sigma\in\Sigma$, the string~$v=\sigma u$ satisfies that~$w_{\ell-2}=0$, and that~$w_0=0$ if and only if~$\sigma=-u_1$. Therefore,~$\chi_{\ell}(v)$ is computed by either~$(3)$ or~$(4)$. Hence, let~$D(u)\triangleq(r,0)$ for~$r\in\bZ_q^{\ell-3}$, and notice that
			\begin{align*}
			\sum_{\sigma\in\bZ_q}\chi_\ell(\sigma u)&=\sum_{\sigma\ne -u_1}\left(T(D(\sigma u)) -\sum_{\tau\ne 0}\frac{P(\sigma+u_1,r,\tau)(\sigma)}{q^{\qnum{\sigma+u_1}{\tau}}} \right)\\
			&\phantom{=}+T(D(-u_1,u))+\sum_{\mu\ne 0}\sum_{\tau\ne 0}\frac{P(\mu,r,\tau)(\mu+(-u_1))}{q^{\qnum{\mu}{\tau} }}\\			
			&=\sum_{\sigma\ne -u_1}T(D(\sigma u)) -\sum_{\sigma\ne -u_1}\sum_{\tau\ne 0}\frac{P(\sigma+u_1,r,\tau)(\sigma)}{q^{\qnum{\sigma+u_1}{\tau} }}\\
			&\phantom{=} +T(D(-u_1,u))+\sum_{\mu\ne -u_1}\sum_{\tau\ne 0}\frac{P(\mu+u_1,r,\tau)(\mu)}{q^{\qnum{\mu+u_1}{\tau} }}\\
			&=\sum_{\sigma\in\bZ_q}T(D(\sigma u))\overset{\eqref{equation:flipSigma}}{=}\sum_{\sigma\in\bZ_q}T(\sigma D( u))=\sum_{\sigma\in\bZ_q}T( D( u)\sigma).
			\end{align*}
		\end{itemize}
		Further, any~$u\in\bZ_{q}^{\ell}$ also satisfies exactly one of the following two cases.
	
		\begin{itemize}
		\item[Case~B1.] $D(u)_0=0$. For any~$\sigma\in\Sigma$, the string~$v=u\sigma$ satisfies that~$w_0=0$, and that~$w_{\ell-2}=0$ if and only if~$\sigma=-u_{\ell-2}$. Therefore,~$\chi_{\ell}(v)$ is computed by either~$(2)$ or~$(4)$. Hence,  let~$D(u)\triangleq (0,r)$ for~$r\in\bZ_q^{\ell-3}$, and notice that
			\begin{align*}
				\sum_{\sigma\in\bZ_q}\chi_\ell(u\sigma)&=\sum_{\sigma\ne -u_{\ell-2}}\left(T(D(u\sigma)) -\sum_{\mu\ne 0}\frac{P(\mu,r,u_{\ell-2}+\sigma)(\mu+u_0)}{q^{\qnum{\mu}{u_{\ell-2}+\sigma}}} \right)\\
				&\phantom{=}+T(D(u,-u_{\ell-2}))+\sum_{\mu\ne 0}\sum_{\tau\ne 0}\frac{P(\mu,r,\tau)(\mu+u_0)}{q^{\qnum{\mu}{\tau}}}\\
				&=\sum_{\sigma\ne -u_{\ell-2}}T(D(u\sigma)) -\sum_{\mu\ne 0}\sum_{\sigma\ne -u_{\ell-2}}\frac{P(\mu,r,u_{\ell-2}+\sigma)(\mu+u_0)}{q^{\qnum{\mu}{u_{\ell-2}+\sigma }}} \\
				&\phantom{=}+T(D(u,-u_{\ell-2}))+\sum_{\mu\ne 0}\sum_{\tau\ne -u_{\ell-2}}\frac{P(\mu,r,u_{\ell-2}+\tau)(\mu+u_0)}{q^{\qnum{\mu}{u_{\ell-2}+\tau } }}\\ 
				&=\sum_{\sigma\in\bZ_q}T(D(u\sigma ))\overset{\eqref{equation:flipSigma}}{=}\sum_{\sigma\in\bZ_q}T( D( u)\sigma)=\sum_{\sigma\in\bZ_q}T(\sigma D( u)).
			\end{align*}
		\item[Case~B2.] $D(u)_0\ne 0$. For any~$\sigma\in\Sigma$, the string~$v=u\sigma$ satisfies that~$w_0\ne 0$, and that~$w_{\ell-2}=0$ if and only if~$\sigma=-u_{\ell-2}$. Therefore,~$\chi_{\ell}(v)$ is computed by either~$(1)$ or~$(3)$. Hence, it follows that
			\begin{align*}
			\sum_{\sigma\in\bZ_q}\chi_\ell(u\sigma)&=\sum_{\sigma\ne -u_{\ell-2}}\left(T(D(u\sigma))+\frac{P(D(u),u_{\ell-2}+\sigma))(u_0)}{q^{\qnum{u_0+u_1}{u_{\ell-2}+\sigma } }} \right)\\
			&\phantom{=}+T(D(u,-u_{\ell-2}))-\sum_{\tau\ne 0}\frac{P(D(u),\tau)(u_0)}{q^{\qnum{u_0+u_1}{\tau} }}\\
			&=\sum_{\sigma\ne -u_{\ell-2}}T(D(u\sigma))+\sum_{\sigma\ne -u_{\ell-2}}\frac{P(D(u),u_{\ell-2}+\sigma))(u_0)}{q^{\qnum{u_0+u_1}{u_{\ell-2}+\sigma} }} \\
			&\phantom{=}+T(D(u,-u_{\ell-2}))-\sum_{\tau\ne -u_{\ell-2}}\frac{P(D(u),u_{\ell-2}+\tau)(u_0)}{q^{\qnum{u_0+u_1}{u_{\ell-2}+\tau } }}\\			
			&=\sum_{\sigma\in\bZ_q}T(D(u\sigma ))\overset{\eqref{equation:flipSigma}}{=}\sum_{\sigma\in\bZ_q}T( D( u)\sigma)=\sum_{\sigma\in\bZ_q}T(\sigma D( u)).
			\end{align*}
	\end{itemize}
	Since any~$u\in\bZ_q^{\ell-1}$ satisfies exactly one of the cases~A1,~A2, and exactly one of the cases~B1,~B2, the claim follows from the correctness of the recursive step. 
\end{IEEEproof}

\begin{lemma}\label{lemma:B_q^ellDistinct}
	All entries of~$\chi_\ell$ are distinct.
\end{lemma}

\begin{IEEEproof}
	Since the output of Algorithm~\ref{algorithm:B_q^ell} is an integer vector, it follows that the minimum absolute difference between the entries of~$T$ is~$2$. Hence, since the additions to the entries of~$T$ are less than~$1$ in absolute value, it follows that all entries with distinct~$D$-values are distinct (i.e., $\chi_\ell(u)\ne \chi_\ell(v)$ for all~$u,v\in\bZ_q^{\ell}$ such that~$D(u)\ne D(v)$). It remains to prove that entries with an identical~$D$-value are distinct. To this end, recall that entries with an identical~$D$-value are of the form~$D^{-1}(w)=\{v_0,\ldots,v_{q-1}\}$ for some~$D$-value~$w$, where~$v_i$ begins with~$i$ for all~$i\in\bZ_q$~\eqref{equation:Dinverse}.
	
	Notice that the additions to the entries of~$T$ are of the form~$\sum_{i=1}^{q^2}\frac{c_i}{q^{i}}$ where~$c_i<q$ for all~$i$. Hence, it is convenient to consider these additions in their fractional $q$-ary expansion. In the sequel, the claim is proven separately for the sets~$D^{-1}(w)$ according to the types~(1), (2), (3), or~(4) of~$w$, as noted in Algorithm~\ref{algorithm:B_q^ell}. For example, $D^{-1}(w)$ is of type~(1) if~$w\in \cA_\ell$, and of type~(3) if~$w_0\ne 0$ and~$w_{\ell-2}=0$. Since the elements in~$D^{-1}(w)=\{v_0,\ldots,v_{q-1}\}$ have distinct leftmost entries~$\{v_{0,0},\ldots,v_{q-1,0} \}$, it follows that --
	\begin{enumerate}
		\item[(1)] the additions~$\frac{P(D(v_i))(v_{i,0})}{q^{\qnum{w_0}{w_{\ell-2}} }}$ to~$T(D(v_i))$ are distinct for any~$i\in\bZ_q$ since~$P(D(v_i))=P(w)$ is a permutation.
		\item[(2)] the additions~$-\sum_{\mu\ne 0}\frac{P(\mu,w'w_{\ell-2})(\mu+v_{i,0})}{q^{\qnum{\mu}{w_{\ell-2}} }}$ to~$T(D(v_i))$ contain the digit $P(\mu,w'w_{\ell-2})(v_{i,0}-\rho)$ in position~$\qnum{\mu}{w_{\ell-2}}$ of the~$q$-ary expansion for any~$\mu\in\bZ_q\setminus\{0\}$, and thus, for any fixed value of~$\mu$, different additions are distinct on this digit.
		\item[(3)] the additions~$-\sum_{\tau\ne 0}\frac{P(w_0,w',\tau)(v_{i,0})}{q^{\qnum{w_0}{\tau} }}$ to~$T(D(v_i))$ contain the digit $P(w_0,w',\tau)(v_{i,0})$ in position~$\qnum{w_0}{\tau}$ of the~$q$-ary expansion for any~$\tau\in\bZ_q\setminus\{0\}$, and thus, for any fixed value of~$\tau$, different additions are distinct on this digit.
		\item[(4)] the additions~$\sum_{\mu\ne 0}\sum_{\tau\ne 0}\frac{P(\mu,w',\tau)(\mu+v_{i,0})}{q^{\qnum{\mu}{\tau} }}$ to~$T(D(v_i))$ are distinct on the~$\qnum{\mu}{\tau}$-th digit for any fixed values of~$\rho$ and~$\tau$.
		
	\end{enumerate}
	Hence, any two additions to $T(D(v_i))$ and $T(D(v_j))$ for~$i,j\in \bZ_q$ such that~$i\ne j$ are distinct, which implies that all entries of~$\chi_\ell$ are distinct.
\end{IEEEproof}

Lemma~\ref{lemma:B_q^ellConstraints} and Lemma~\ref{lemma:B_q^ellDistinct} imply that the output of Algorithm~\ref{algorithm:B_q^ell} is indeed a feasible matrix. It remains to prove that the outputs of Algorithm~\ref{algorithm:B_q^ell} for two distinct information vectors correspond to distinct feasible permutations, i.e., that Algorithm~\ref{algorithm:B_q^ell} is injective.

\begin{lemma}
	If~$s$ and~$t$ are distinct information vectors in~$J_q^\ell$ such that $B_q^\ell(s)=q^{q^2}\chi_\ell^s,B_q^\ell(t)=q^{q^2}\chi_\ell^{t}$ and $\chi_\ell^s\vDash \pi_s,\chi_\ell^{t}\vDash \pi_{t}$ for some permutations~$\pi_s$ and~$\pi_{t}$, then $\pi_s\ne \pi_{t}$.
\end{lemma}

\begin{IEEEproof}
	Since in every stage~$i$ of the algorithm, the additions to the entries of the respective~$T$ are less than half of the minimum absolute distance between the entries of~$T$, it follows that Algorithm~\ref{algorithm:B_q^ell} preserves the order among the distinct homomorphic pre-images of~$D$. That is, for any positive integers~$i$ and~$j$, and any~$u,v\in\bZ_q^i$, if~$\chi_i(u)>\chi_i(v)$, then for all~$u',v'\in\bZ_q^{i+j}$ such that~$D^j(u')=u$ and~$D^j(v')=v$, we have that~$\chi_{i+j}(u')>\chi_{i+j}(v')$.
	
	If~$\ell=2$, the claim follows from Lemma~\ref{lemma:A_qDistinct}. Otherwise, denote~$s=(s',P_3^s,\ldots,P_\ell^s)$ and~$t=(t',P_3^t,\ldots,P_\ell^t)$. If~$s'\ne t'$ then in stage~$\ell=3$ of the algorithm, by Lemma~\ref{lemma:A_qDistinct} there exist distinct~$u$ and~$v$ in~$\bZ_q^2$ on whom~$\chi_2^s$ and~$\chi_2^t$ disagree, i.e., $\chi_2^s(u)>\chi_2^s(v)$ and $\chi_2^t(u)<\chi_2^t(v)$. It follows that~$\chi_3^s$ and~$\chi_3^t$ disagree on any~$u',v'\in\bZ_q^3$ such that~$D(u')=u$ and~$D(v')=v$. If~$s'=t'$, then assume that~$P_i^s\ne P_i^t$ for some~$i\in \{3,\ldots,\ell \}$, which implies that there exists~$u\in\cA_i$ that is mapped by~$P_i^s$ and~$P_i^t$ to distinct permutations in~$S_q$. Hence, the pre-images $D^{-1}(u)$ are ordered differently in $\chi_i$. Further, according to the above discussion, all pre-images of~$D^{-1}(u)$ are ordered differently in all~$\chi_j$ for~$j>i$, and the claim follows.
\end{IEEEproof}

The above discussion, together with Corollary~\ref{corollary:numFeasible_ell=2}, implies the following lower bound on the number of feasible permutations. By the definition of~$\cA_i$~\eqref{equation:DefJ_qell} we have that its size is $q^{i-1}-2q^{i-2}+q^{i-3}$. Hence, the size of~$\cP_i$ is~$(q!)^{q^{i-1}-2q^{i-2}+q^{i-3}}$, and the following corollary is immediate.

\begin{corollary}\label{corollary:numOfFeasibleAnyEll}
	For any~$q\ge 3$ and~$\ell\ge 2$, 
	\[
	\cF_{q,\ell}\ge f_{3,2}\cdot\prod_{j=4}^{q}\left(j!\cdot {j^2-j+1\choose j} \right)
	\cdot\left(\prod_{i=3}^{\ell}(q!)^{q^{i-1}-2q^{i-2}+q^{i-3}}\right).
	\]
\end{corollary}

To estimate the contribution of Algorithm~\ref{algorithm:B_q^ell}, one may take either~$q$ or~$\ell$ to infinity. The proof of the following claim is given in Appendix~\ref{section:omittedProofs}.

\begin{lemma}\label{lemma:Alg2q2Inf}
	For any constant~$\ell\ge 3$, $\lim_{q\to\infty} R_{q,\ell}\ge \frac{1}{\ell}$.
\end{lemma}

When applications in DNA storage are discussed, it is natural to keep~$q$ a constant. However, the rate of the set of permutations which is given by Algorithm~\ref{algorithm:B_q^ell} goes to zero as~$\ell$ tends to infinity. Hence, for finite values of~$q$ and~$\ell$, lower bounds for the corresponding the rates are given in the following table. Due to computational restrictions, the value of entry~$(q,\ell)$  is $\frac{\log(q!)(q-1)(q^{\ell-2}-1)}{\ell q^\ell \log(q)}$, which is a lower bound on~$R_{q,\ell}$.

\begin{center}
	\begin{tabular}{|c|c|c|c|c|c|c|c|c|c|}
	\hline
	\diagbox{$q$}{$\ell$} & 3      &   4    &   5    &   6    &   7    &   8    &   9    &   10   \\ \hline
				3              & 0.0805 & 0.0805 & 0.0698 & 0.0597 & 0.0516 & 0.0452 & 0.0403 & 0.0362 \\ \hline
				4              & 0.1075 & 0.1007 & 0.0846 & 0.0714 & 0.0613 & 0.0537 & 0.0478 & 0.0430 \\ \hline
				5              & 0.1269 & 0.1142 & 0.0944 & 0.0792 & 0.0680 & 0.0595 & 0.0529 & 0.0476 \\ \hline
				6              & 0.1417 & 0.124  & 0.1015 & 0.0849 & 0.0728 & 0.0637 & 0.0567 & 0.0510 \\ \hline
				7              & 0.1533 & 0.1314 & 0.1070 & 0.0894 & 0.0766 & 0.0671 & 0.0596 & 0.0536 \\ \hline
				8              & 0.1627 & 0.1373 & 0.1113 & 0.0929 & 0.0797 & 0.0697 & 0.062  & 0.0558 \\ \hline
				9              & 0.1705 & 0.1421 & 0.1149 & 0.0959 & 0.0822 & 0.0719 & 0.0639 & 0.0575 \\ \hline
				10             & 0.1771 & 0.1461 & 0.1180 & 0.0984 & 0.0843 & 0.0738 & 0.0656 & 0.0590 \\ \hline
\end{tabular}
\end{center}

\section{String Length for Feasible Permutations}\label{section:length}
In order to obtain a practical rank-modulation scheme from a given set of feasible permutations, it is essential to estimate the minimal lengths of the corresponding strings. To this end, for a feasible permutation~$\pi\in S_{q,\ell}$, let $\len (\pi)\triangleq\min\{n ~\vert~ \exists x\in\Sigma^n, \pi\vDash x \}$, and for a set~$S\subseteq S_{q,\ell}$ of feasible permutations, let $\len (S)\triangleq \max \{\len(\pi) \}_{\pi\in S}$. In this subsection an upper bound for~$\len(T_{q,\ell})$ is given, where~$T_{q,\ell}$ is the set of permutations which are given by Algorithm~\ref{algorithm:B_q^ell} for the parameters~$q$ and~$\ell$. It will be evident that for a constant~$q$, the permutations in~$T_{q,\ell}$ have corresponding strings whose lengths are polynomial in~$q^\ell$. The proofs in this subsection rely on the following simple lemma.

\begin{lemma}\label{lemma:minimalLengthSumOfEntries}
	If $\pi\in S_{q,\ell}$ is a feasible permutation and~$\chi\in \bN^{q^\ell}$ is a feasible vector such that $\pi\vDash \chi$, then~$\len(\pi)\le \sum_{v\in\Sigma^{q^\ell}}\chi(v)$.
\end{lemma}

\begin{IEEEproof}
	Since~$\chi$ is feasible, it follows that~$\sum_{\sigma \in\Sigma}\chi(u\sigma)=\sum_{\sigma \in\Sigma}\chi(\sigma u)$ for all~$u\in\bZ_{q^{\ell-1}}$. Hence, there exists a closed path in~$G_q^\ell$ that traverses edge~$v$ for~$\chi(v)$ times for all~$v\in\Sigma^\ell$. Clearly, the length of the corresponding string~$x$ is~$\sum_{v\in\Sigma^\ell}\chi(v)$, and $x\vDash\pi$.
\end{IEEEproof}

For the set~$T_{q,2}$ let~$c_q\triangleq\min_\Phi \{\max(\chi)\vert \chi\in\Phi \}$, where~$\Phi$ ranges over all sets of feasible vectors in~$\bN^{q^2}$ that contain a unique feasible vector for every permutation in~$T_{q,2}$. The following lemma provides a bound for~$c_q$ as a function of~$c_3$. Using a computer program, it was discovered that~$c_3 \le 16$.

\begin{lemma}\label{lemma:entryBoundforL=2}
	$c_q\le 2^{q-3}\cdot\frac{q!}{6}\cdot\frac{(q+1)!}{24}\cdot c_3$.
\end{lemma}

\begin{IEEEproof}
	It is evident from Algorithm~\ref{algorithm:A_q} that an entry in~$\chi$ is at most~$c_{q-1}(q+1)+1$ if it is not in the top row or leftmost column, and $c_{q-1}(q+1)+q+1$ if it is. Hence,
	\begin{align*}
		c_q&\le q(c_{q-1}(q+1)+q+1)=q(q+1)(c_{q-1}+1)\\
		& \le 2q(q+1)c_{q-1}\le \ldots \le 2^{q-3}\cdot\frac{q!}{6}\cdot\frac{(q+1)!}{24}\cdot c_3.
	\end{align*}
\end{IEEEproof}

In turn, Lemma~\ref{lemma:minimalLengthSumOfEntries} and Lemma~\ref{lemma:entryBoundforL=2} provide the following.

\begin{corollary}\label{corollary:lengthL=2}
	$\len(T_{q,2})\le q^2\cdot 2^{q-3}\cdot\frac{q!}{6}\cdot\frac{(q+1)!}{24}\cdot c_3$.
\end{corollary}

Let~$c_{q,\ell}\triangleq\min_\Phi\{\max(\chi)\vert \chi \in\Phi \}$, where~$\Phi$ ranges over all sets of vectors in~$\bN^{q^\ell}$ that contain a unique feasible vector for every permutation in~$T_{q,\ell}$. For the next lemma, notice that $c_{q,2}=c_q$ by definition.

\begin{lemma}\label{lemma:minimalLength}
	$c_{q,\ell}\le c_q (3q^{q^2})^{\ell-2}$.
\end{lemma}

\begin{IEEEproof}
	Since the additions to the entries of~$T$ in Algorithm~\ref{algorithm:B_q^ell} are of absolute value at most~1, it follows from Line~\ref{algorithm2Line:multiply}, Line~\ref{algorithm2Line:define} and Line~\ref{algorithm2Line:return} of Algorithm~\ref{algorithm:B_q^ell} that~$c_{q,\ell}\le (2c_{q,\ell-1}+1)q^{q^2}\le c_{q,\ell-1}(3q^{q^2})$. Solving this recursion relation proves the claim.
\end{IEEEproof}

The proof of the following corollary is immediate from Lemma~\ref{lemma:minimalLength}.

\begin{corollary}\label{corollary:minimalLength}
	$\len(T_{q,\ell})\le c_3\cdot\frac{2^{q-3}\cdot\frac{q!}{6}\cdot\frac{(q+1)!}{24}}{2q^2} \cdot 3^{\ell-2}\left(q^\ell \right)^{q^2+1}$.
\end{corollary}

It is evident from Corollary~\ref{corollary:minimalLength} that Algorithm~\ref{algorithm:B_q^ell} provides permutations whose corresponding strings are of lengths that is polynomial in~$q^\ell$ for a fixed value of~$q$. However, the constants that are involves in this bound, including the constant in the exponent, are rather large even for small values of~$q$.

\section{Discussion}\label{section:discussion}

In this paper the question of feasibility of permutations was addressed. Our contributions include an upper bound on the number of feasible permutations, a linear programming algorithm for deciding the feasibility of a permutation, and a recursive algorithm for explicit construction of a large feasible set. Further, the length of the strings which correspond to permutations in this set was shown to be polynomial in~$q^\ell$ for a fixed~$q$. In addition, in Appendix~\ref{section:minimumDistance} it is shown how feasible permutations of minimum Kendall-$\tau$ distance can be produced by pre-coding the information vector in Algorithm~\ref{algorithm:B_q^ell}. However, the resulting distance is rather low.

The most prominent directions for future research seem to be studying the values of~$R_{q,\ell}$ and~$\len (\cF_{q,\ell})$, and providing better constructions in terms of rate, distance, and length. Furthermore, alternative combinatorial models for DNA storage should be studied. For example, one may group the entries of the profile vector by sum or by cyclic equivalence, and study the achievable rates of applying rank modulation schemes on the resulting sets of vectors.

\bibliographystyle{IEEEtranS}
\bibliography{allbib}

\appendices

\section{Omitted Proofs}\label{appendix:OmittedProofs}
\label{section:omittedProofs}
\begin{IEEEproof} (Proof of Lemma~\ref{lemma:Rq2})
  According to Eq.~\eqref{equation:FqlRql} and Corollary~\ref{corollary:numFeasible_ell=2}, it follows that for any~$q\ge 4$, and any real number $\alpha>1$,
  
  \begin{align*}
    R_{q,2}&\ge\frac{1}{\log(q^2!)}\log\parenv{f_{3,2}\cdot\prod_{j=4}^{q}j!\cdot{j^2-j+1\choose j}} \\
    &\ge \frac{1}{\log(q^2!)}\parenv{\sum_{i\geq 1}\sum_{j=\floor{q/\alpha^i}+1}^{\floor{q/\alpha^{i-1}}} \left(\log(j!)+\log{j^2-j+1\choose j}\right)}\\
    &\ge \frac{1}{\log(q^2!)}\sum_{i\geq 1}\parenv{\parenv{\bigfloor{\frac{q}{\alpha^{i-1}}} - \bigfloor{\frac{q}{\alpha^{i}}}} \left(\log\parenv{\bigfloor{\frac{q}{\alpha^i}}!}+\log{\floor{\frac{q}{\alpha^i}}^2-\floor{\frac{q}{\alpha^i}}+1\choose \floor{\frac{q}{\alpha^i}}}\right)},
  \end{align*}
  By using the simple lower bound~${s\choose t}\ge
  (\frac{s}{t})^t$, the
  identity~$\lim_{m\to\infty}\frac{\log(m!)}{m\log m}=1$, and by omitting the rounding operation when~$q$ is large enough, it follows
  that
  \begin{align*}
    \lim_{q\to\infty}R_{q,2}&\ge \lim_{q\to\infty}\sum_{i\geq 1}\frac{1}{2q^2\log q }\parenv{\frac{2q^2(\alpha-1)}{\alpha^{2i}}\log \frac{q}{\alpha^i}}\\
    &= \sum_{i\geq 1}\lim_{q\to\infty}\frac{1}{2q^2\log q }\parenv{\frac{2q^2(\alpha-1)}{\alpha^{2i}}\log \frac{q}{\alpha^i}} \\
    &= \sum_{i\geq 1}\frac{\alpha-1}{\alpha^{2i}}=\frac{1}{1+\alpha},
  \end{align*}
  with a simple application of dominated convergence. Since this holds for
  every $\alpha>1$,
  \[ \lim_{q\to\infty}R_{q,2} \geq \lim_{\alpha\to 1}\frac{1}{1+\alpha}=\frac{1}{2}.\]
\end{IEEEproof}

\begin{IEEEproof}(of Lemma~\ref{lemma:Alg2q2Inf})
	According to Corollary~\ref{corollary:numOfFeasibleAnyEll}, it follows that for any~$q\ge 3$ and~$\ell\ge 3$,
	\begin{align*}
		R_{q,\ell}\ge \frac{\log\left( f_{3,2} \prod_{j=4}^q \left(j!{j^2-j+1\choose j}\right)\cdot \left(\prod_{i=3}^{\ell} (q!)^{q^{i-1}-2q^{i-2}+q^{i-3} } \right) \right)}{\log(q^\ell!)},
	\end{align*}
	and hence,
	\begin{align*}
		\lim_{q\to \infty}R_{q,\ell}&=
		\lim_{q\to \infty} \frac{\sum_{i=3}^{\ell}(q^{i-1}-2q^{i-2}+q^{i-3})q\log q}{\ell q^\ell \log q}=\frac{1}{\ell}
	\end{align*}
\end{IEEEproof}

\begin{lemma}\label{lemma:TransitionMatrix}
	For a given vector positive vector~$s\in\bR^{q^\ell}$ whose sum of entries is~$1$, the matrix~$M_s$~\eqref{equation:TransitionMatrix} is a transition matrix of a Markov chain on a DeBruijn graph, whose stationary distribution is~$s$.
\end{lemma}
\begin{IEEEproof}
	For given~$v\in\Sigma^{\ell-1}$, the sum of row~$\tau v$ of~$M_s$ for any~$\tau\in\Sigma$ is
	\begin{align*}
		\sum_{\sigma \in \Sigma}\frac{s(v\sigma)}{\sum_{\rho\in\Sigma}s(v\rho)}=\frac{\sum_{\sigma\in\Sigma}s(v\sigma)}{\sum_{\rho\in\Sigma}s(v\rho)}=1,
	\end{align*}
	and hence~$M_s$ is a transition matrix. Further, for any~$v\in\Sigma^{\ell-1}$ and~$\sigma\in\Sigma$, the~$v\sigma$-th entry of~$sM_s$ equals
	\begin{align*}
		\left(sM_s\right)_{v\sigma}&=\sum_{\rho\in\Sigma}s(\rho v)\cdot (M_s)_{\rho v,v\sigma}\\
		&= \sum_{\rho\in \Sigma}\frac{s(\rho v)\cdot s(v\sigma)}{\sum_{\tau\in\Sigma}s(v\tau)}\\
		&= s(v\sigma)\cdot\frac{\Sigma_{\rho\in\Sigma}s(\rho v)}{\Sigma_{\tau\in\Sigma}s(v\tau )}=s(v\sigma),
	\end{align*}
	where the last equality follows from the flow-conservation of~$s$. Hence,~$s$ is the stationary distribution of~$M_s$.
\end{IEEEproof}

\section{Minimum Distance by the Kendall~$\tau$ metric}
\label{section:minimumDistance}
Recall that the purpose of applying a rank-modulation scheme for the DNA storage channel is to endure small errors in the profile vector which is given at the output of the channel. Clearly, any pattern of errors that does not cause two entries in the profile vector to surpass one another is correctable. However, if the channel is restricted to a rank-modulation code by the Kendall-$\tau$ distance (see below), error resilience is improved at the cost of lower information rate. In particular, if the channel is restricted to permutations that are of distance at least~$2t+1$ apart, then any error pattern of at most~$t$ adjacent transpositions is correctable. Fortunately, the structure of the information vectors in Algorithm~\ref{algorithm:A_q} and Algorithm~\ref{algorithm:B_q^ell} allows rank-modulation codes to be incorporated conveniently. However, the resulting distances are rather low for small values of~$q$.

First, it is worth noting that each recursive step of Algorithm~\ref{algorithm:A_q} relies on \textit{interleaving} a new set of strings, which contain the newly added symbol~$q-1$, into a permutation on~$\bZ_{q-1}^2$. The permutation among these added strings, and the particular way by which the interleaving is made, are determined by the information vector. In what follows, this intuition is formalized, and the resulting minimum distance is discussed.

\begin{definition}\label{definition:KendallTau}
	The Kendall-$\tau$ distance~$d_\tau$ between two strings is the minimum number of adjacent transpositions that can be applied on one to obtain the other. 
\end{definition}

Although Definition~\ref{definition:KendallTau} is usually applied over permutations, i.e., for strings which contain all symbols of the alphabet with no repetitions, it may also be applied over ordinary strings.

\begin{example}\label{example:KendallTauBinary}
	$d_\tau(10010,00110)=2$.
\end{example}

For disjoint sets of symbols~$A$ and~$B$ let~$C_A\subseteq S_A$ and~$C_B\subseteq S_B$ be codes of minimum Kendall-$\tau$ distances~$d_A$ and~$d_B$, respectively, and let~$D\subseteq \{0,1 \}^{|A|+|B|}$ be of constant Hamming weight~$|A|$ and minimum Kendall-$\tau$ distance~$d_D$. Define the operator~$*_D$ as 
\begin{align}\label{equation:starD}
C_A*_D C_B\triangleq \{\pi\in S_{A\cup B}~:~\pi\vert_A\in C_A,\pi\vert_B\in C_B, f(\pi)\in D \},
\end{align}
where~$\pi\vert_A$ (resp.~$\pi\vert_B$) denotes the result of deleting all symbols that are not in~$A$ (resp.~$\pi\vert_B$) from~$\pi$, and 
\begin{align*}
f:S_{A\cup B}&\to \{0,1\}^{|A|+|B|}\\
f(\pi)_i&=\begin{cases}
1 & \pi_i\in A\\
0 & \pi_i\in B
\end{cases}.
\end{align*}

\begin{lemma}\label{lemma:starDdistance}
	The minimum distance~$d_\tau(\cC)$ of~$\cC\triangleq C_A *_D C_B$ is at least~$\min \{d_A,d_B,d_D \}$.
\end{lemma}

\begin{IEEEproof}
	It is readily verified that the mapping~$g:S_{A\cup B}\to S_A\times S_B\times \{0,1\}^{|A|+|B|}$, $\pi\mapsto(\pi\vert_A,\pi\vert_B,f(\pi))$ is injective. Further, notice that for an adjacent transposition~$e$, if it affects two elements from~$A$ then~$g(\pi)$ and~$g(e(\pi))$ differ only on the~$S_A$ component. Similarly, if~$e$ affects two elements from~$B$ then~$g(\pi)$ and~$g(e(\pi))$ differ only on the~$S_B$ component, and if it affects one element from~$A$ and one from~$B$ then~$g(\pi)$ and~$g(e(\pi))$ differ only on the~$\{0,1\}^{|A|+|B|}$ component. Hence, for~$\pi_1$ and~$\pi_2$ in~$\cC$, 
	\begin{align*}
	d_\tau(\pi_1,\pi_2)=d_\tau(\pi_1\vert_A,\pi_2\vert_A )+d_\tau(\pi_1\vert_B,\pi_2\vert_B )+d_\tau(f(\pi_1),f(\pi_2)),
	\end{align*}
	from which the claim follows.
\end{IEEEproof}

To obtain a minimum distance guarantee at the output of Algorithm~\ref{algorithm:A_q}, let~$C_3$ be a rank-modulation code in~$T_{3,2}$ (where the notation~$T_{q,\ell}$ is as in Section~\ref{section:length}) of minimum Kendall-$\tau$ distance~$d_3$, for all~$i\in\{4,\ldots,q\}$ let~$C_i$ be a rank-modulation code in~$S_i$ of minimum Kendall-$\tau$ distance~$d_i$, and let~$B_i\subseteq \{0,1\}^{i^2-i+1}$ be a binary code of constant Hamming weight~$i$ and minimum Kendall-$\tau$ distance~$t_i$. Replace the information set~$I_q$~\eqref{equation:I_q} by the set
\begin{align}\label{equation:I_q'}
I_q'\triangleq C_3\times (C_4\times B_4)\times\cdots\times (C_q\times  B_q).
\end{align}

\begin{lemma}\label{lemma:I_q'}
	If the information vectors in Algorithm~\ref{algorithm:A_q} are taken from~$I_q'$ rather than from~$I_q$, then the minimum distance of the resulting permutations is at least~$\min(\{d_i \}_{i=3}^q\cup\{t_i \}_{i=4}^q)$.
\end{lemma}

\begin{IEEEproof}
	Let~$T_{q,2}'$ be the permutations which result from Algorithm~\ref{algorithm:A_q} when applied over information vectors from~$I_q'$. The claim is proved by using induction on~$q$. It follows from the definition of the algorithm that~$T_{3,2}=C_3$, which proves the claim for~$q=3$. Assume that $d_\tau(T_{q-1,2}')\ge \min(\{d_i\}_{i=3}^{q-1}\cup \{t_i \}_{i=4}^{q-1} )$. Notice that for~$q\ge 4$, the set~$T_{q,2}'$ is given by considering $T_{q-1,2}' *_{B_q} C_q$, replacing the elements of~$[q]$ by the strings which correspond to~$y_0,\ldots,y_{q-1}$, and adding the strings which corresponds to~$y_0-(q-2)\varepsilon$ and to~$y_i-\varepsilon$ for~$1\le i\le q-2$, at fixed positions. Since the addition of the latter strings may only increase the minimum distance in comparison with~$T_{q-1,2}' *_{B_q} C_q$, it follows from Lemma~\ref{lemma:starDdistance} that
	\begin{align*}
	d_\tau(T_{q,2}')\ge d_\tau(T_{q-1,2}' *_{B_q} C_q)\ge \min\{d_\tau(T_{q-1,2}'), d_\tau(B_q), d_\tau(C_q) \}.
	\end{align*}
	Hence, since the induction hypothesis implies that~$d(T_{q-1,2}')\ge \min(\{d_i\}_{i=3}^{q-1}\cup \{t_i \}_{i=4}^{q-1} )$, since~$d_\tau(B_q)=t_q$, and since~$d_\tau(C_q)=d_q$, the result follows.
\end{IEEEproof}

If follows from~Lemma~\ref{lemma:I_q'} that by pre-coding the information vectors into~$I_q'$, a non-trivial minimum distance guarantee is obtained. Notice that codes of size one, whose minimum distance is infinite, can be used as either of the~$C_i$-s or~$B_i$-s in~\eqref{equation:I_q'} to increase the minimum distance of the resulting code.

To incorporate similar approach to Algorithm~\ref{algorithm:B_q^ell}, notice that each recursive step of it relies on \textit{splitting} any string~$w\in\bZ_q^{\ell-1}$ to~$q$ strings of the form~$D^{-1}(w)\subseteq \bZ_q^\ell$. The order between strings in~$\bZ_{q}^\ell$ with different~$D$-preimage is consistent with that of their~$D$-ancestors, whereas the order between strings in~$\bZ_{q}^\ell$ with identical~$D$-preimage is determined by information vector. To state this intuition in the spirit of~\eqref{equation:starD}, the following definitions are given. Let~$A$ and~$B$ be disjoint sets, and for~$u\in S_A$,~$v\in S_B$, and~$b\in B$, let~$h(u,b,v)$ be the permutation on~$A\cup (B\setminus \{ b\})$ which results from replacing the occurrence of~$b$ in~$v$ by~$u$.

\begin{example}
	If $A=\{1,2,3\},B=\{4,5,6\}$, and $u=(3,1,2)$, $v=(4,5,6)$, then $h(u,4,v)=(3,1,2,5,6)$.
\end{example}

For~$C_A\subseteq S_A$, $C_B\subseteq S_B$, and~$b\in B$, let~$h(C_A,b,C_B)\triangleq \{h(u,b,v)\vert u\in C_A,v\in C_B \}$. In an inductive manner, for pairwise disjoint sets~$\{A_1,\ldots,A_t,B \}$ and distinct~$b_1,\ldots,b_t$, let 
\begin{align*}
h((C_{A_i})_{i=1}^t,(b_i)_{i=1}^t,C_B )\triangleq h(C_{A_t},b_t,h((C_{A_i})_{i=1}^{t-1},(b_i)_{i=1}^{t-1},C_B ) ),
\end{align*}
which reads as replacing~$b_i$ in every codeword of~$C_B$ by codewords of~$C_{A_i}$, for every~$i\in [t]$.

\begin{lemma}\label{lemma:minDist-h}
	Let~$\{A_1,\ldots,A_{|B|},B\}$ be disjoint sets such that~$|A_i|\triangleq q$ for all~$i$, let~$C_{A_i}\subseteq S_{A_i}$ be a code of minimum distance~$d_i$ for all~$i$, and let~$C_B\subseteq S_B$ be a code of minimum distance~$d_B$. The minimum distance~$d_\tau(\cH)$ of~$\cH\triangleq h((C_{A_i})_{i=1}^{|B|},(b_i)_{i=1}^{|B|},C_B )$ is at least~$\min(\{d_i \}_{i=1}^{|B|}\cup \{q^2d_B \})$.
\end{lemma}

\begin{IEEEproof}
	For~$\pi\in \cH$ and~$A\triangleq \cup_{i=1}^{|B|}A_i$, let~$r: S_{A}\to B^{q|B|}$ be such that $r(\pi)_i$ equals~$b_j$ for the unique~$j\in\{1,\ldots,|B|\}$ for which~$\pi_i\in A_j $; intuitively, the function~$r$ identifies the index $j\in\{1,\ldots,|B| \}$ of the set~$A_j$ from which each symbol~$\pi_i$ in~$\pi$ is taken, and places~$b_j$ instead of~$\pi_i$. It is readily verified that the function~$g:S_A\to S_{A_1}\times \ldots\times S_{A_{|B|}}\times B^{q|B|}$, $\pi\mapsto(\pi\vert_{A_1},\ldots,\pi\vert_{A_{|B|}},r(\pi))$ is injective. Further, notice that for an adjacent transposition~$e$, if it affects two elements from some~$A_i$, then~$g(\pi)$ and~$g(e(\pi))$ differ only on the~$S_{A_i}$ component. On the other hand, if~$e$ affects two elements from distinct~$A_i$ and~$A_j$, then~$g(\pi)$ and~$g(e(\pi))$ differ only on the~$B^{q|B|}$ component. 
	
	Hence, let~$\pi_1$ and~$\pi_2$ be codewords in~$\cH$, and let~$c_1$ and~$c_2$ be the corresponding codewords from~$C_B$ from which~$\pi_1$ and~$\pi_2$ were generated. The minimal set of adjacent transpositions which differs~$\pi_1$ from~$\pi_2$ contains~$q^2$ transpositions for each transposition which differs~$c_1$ from~$c_2$. Furthermore, it contains a unique transposition on~$A_i$ for each transposition which differs~$\pi_1\vert_{A_i}$ from~$\pi_2\vert _{A_i}$, for each~$i$. Therefore,
	\begin{align*}
	d_\tau(\pi_1,\pi_2)=\sum_{i=1}^{|B|}d_\tau(\pi_1\vert_{A_i},\pi_2\vert_{A_i}) + q^2d_\tau(c_1,c_2),
	\end{align*}
	which proves the claim.
\end{IEEEproof}

Similar to~\eqref{equation:I_q'}, to obtain a rank-modulation code at the output of Algorithm~\ref{algorithm:B_q^ell}, a different set of information vectors is used. Recall that the information vector~$J_q^\ell$ of Algorithm~\ref{algorithm:B_q^ell} \eqref{equation:DefJ_qell} consists of mappings from sets of strings~$\cA_i$ into~$S_{\bZ_q}$. Due to Lemma~\ref{lemma:minDist-h}, the set~$S_{\bZ_q}$ can be replaced in~$J_q^\ell$ by some rank-modulation code. However, due to the~$q^2$ factor in the recursive term of Lemma~\ref{lemma:minDist-h}, it suffices to replace~$S_{\bZ_q}$ only in the rightmost entry of~$J_q^\ell$. To this end, let~$C\subseteq S_{\bZ_q}$ be a rank-modulation code of minimum distance~$d$, and let
\begin{align*}
P_\ell^C&\triangleq \{P\vert P:\cA_i\to C \},\mbox{ and}\\
J_{q,C}^\ell &\triangleq I_q\times \cP_3\times \ldots\times \cP_{\ell-1}\times P_\ell^C.
\end{align*}

\begin{lemma}\label{lemma:Algoritm2MinimumDistance}
	For~$\ell\ge 3$, if the information vectors in Algorithm~\ref{algorithm:B_q^ell} are taken from~$J_{q,C}^\ell$ rather than from~$J_q^\ell$, then the minimum distance of the resulting permutations is at least~$d$.
\end{lemma}

\begin{IEEEproof}
	Let~$T_{q,\ell}'$ be the permutations which result from Algorithm~\ref{algorithm:B_q^ell} when applied over information vectors from~$J_{q,C}^\ell$. The claim is proved using induction on~$\ell$. For~$\ell=3$, the set~$T_{q,3}'$ is given by splitting permutations on~$\bZ_q^2$. Each string~$w\in\cA_3\subseteq\bZ_{q}^2$ is replaced by a permutation~$P(w)$ on~$D^{-1}(w)$ from\footnote{Note that the~$q$ elements of~$D^{-1}(w)$ are arranged according to a codeword the code~$C\subseteq S_q$, and hence it may be seen as a code~$C_{D^{-1}(w)}$ on~$D^{-1}(w)$ of identical minimum distance~$d$.} the code~$C_{D^{-1}(w)}$. Subsequently, strings $u\in\bZ_{q}^2\setminus \cA_3$ are replaced with permutations that depend only on the values of~$\{P(w)\vert w\in \cA_3\}$. Hence, since removing elements may only decrease the minimum distance, it suffices to bound the minimum distance of the permutations of~$\cup_{w\in \cA_3} D^{-1}(w)$. The latter is given as
	\begin{align*}
	h\left( (C_W)_{W\in D^{-1}(\cA_3)}, (a)_{a\in\cA_3}, T_{q,2} \right)\mbox{, where}\\
	D^{-1}(\cA_3) \triangleq \{D^{-1}(w)\vert w\in \cA_3\}.
	\end{align*}
	Therefore, according to Lemma~\ref{lemma:minDist-h}, it follows that $d_\tau(T_{q,3}')\ge \min \{d,q^2\cdot d_\tau(T_{q,2}) \}=d$.
	
	For any~$\ell\ge 4$, similar arguments show that it suffices to bound the minimum distance of
	\begin{align*}
	h\left( (C_W)_{W\in D^{-1}(\cA_\ell)}, (a)_{a\in\cA_\ell}, T_{q,\ell-1}' \right)\mbox{, where}\\
	D^{-1}(\cA_\ell) \triangleq \{D^{-1}(w)\vert w\in \cA_\ell\},
	\end{align*}
	and by again by Lemma~\ref{lemma:minDist-h}, it follows that $d_\tau(T_{q,\ell}')\ge \min \{d,q^2\cdot d_\tau(T_{q,\ell-1}') \}=d$.
\end{IEEEproof}
\end{document}